\documentclass[epj]{svjour}
\usepackage{amsmath,amssymb,latexsym,amsfonts,feynmf}
\begin{document}
\title{On the trees of quantum fields}
\author{Christian Brouder}
\institute{Laboratoire de Min\'eralogie-Cristallographie, CNRS UMR7590,
 Universit\'es Paris 6, Paris 7, IPGP, Case 115, 4 place Jussieu\\
  \small 75252 Paris {\sc cedex} 05, France\\
  \small \tt brouder@lmcp.jussieu.fr}
\date{\today}
\abstract{The solution of some equations involving functional derivatives
is given as a series indexed by planar binary trees. The terms of the
series are given by an explicit recursive formula. Some algebraic properties
of these series are investigated. Several examples are treated in the
case of quantum electrodynamics: the complete fermion and photon propagators,
the two-body Green function, and the one-body Green function
in the presence of an external
source, the complete vacuum polarization, electron self-energy and
irreducible vertex.
\PACS{{03.70.+k}{Theory of quantized fields}   \and
      {11.15.Bt}{General properties of perturbation theory} \and
      {12.20.-m}{Quantum electrodynamics}
     } 
} 
\maketitle

\begin{fmffile}{ChBtree}
\unitlength=1mm


\newcommand{\tunn}{\parbox{3mm}{
\begin{fmfchar*}(3,3)
  \fmfforce{(0.5w,0.5h)}{i1}
  \fmfv{decor.shape=circle,decor.filled=1,decor.size=20}{i1}
\end{fmfchar*}}}

\newcommand{\tunb}{\parbox{3mm}{
\begin{fmfchar*}(3,3)
  \fmfforce{(0.5w,0.5h)}{i1}
  \fmfv{decor.shape=circle,decor.filled=0,decor.size=20}{i1}
\end{fmfchar*}}}



\newcommand{\ttroisn}{
\parbox{5mm}{
\begin{fmfchar*}(5,4)
  \fmfforce{(0.5w,0.1h)}{i1}
  \fmfforce{(0.1w,0.9h)}{o1}
  \fmfforce{(0.9w,0.9h)}{o2}
  \fmf{vanilla}{i1,o1}
  \fmf{vanilla}{i1,o2}
  \fmfv{decor.shape=circle,decor.filled=1,decor.size=20}{i1}
  \fmfv{decor.shape=circle,decor.filled=0,decor.size=20}{o1}
  \fmfv{decor.shape=circle,decor.filled=1,decor.size=20}{o2}
\end{fmfchar*}
}}

\newcommand{\ttroisnn}{
\parbox{5mm}{
\begin{fmfchar*}(5,4)
  \fmfforce{(0.5w,0.1h)}{i1}
  \fmfforce{(0.1w,0.9h)}{o1}
  \fmfforce{(0.9w,0.9h)}{o2}
  \fmf{vanilla}{i1,o1}
  \fmf{vanilla}{i1,o2}
  \fmfv{decor.shape=circle,decor.filled=1,decor.size=20}{i1}
  \fmfv{decor.shape=circle,decor.filled=1,decor.size=20}{o1}
  \fmfv{decor.shape=circle,decor.filled=1,decor.size=20}{o2}
\end{fmfchar*}
}}

\newcommand{\ttroisb}{
\parbox{5mm}{
\begin{fmfchar*}(5,4)
  \fmfforce{(0.5w,0.1h)}{i1}
  \fmfforce{(0.1w,0.9h)}{o1}
  \fmfforce{(0.9w,0.9h)}{o2}
  \fmf{vanilla}{i1,o1}
  \fmf{vanilla}{i1,o2}
  \fmfv{decor.shape=circle,decor.filled=0,decor.size=20}{i1}
  \fmfv{decor.shape=circle,decor.filled=0,decor.size=20}{o1}
  \fmfv{decor.shape=circle,decor.filled=1,decor.size=20}{o2}
\end{fmfchar*}
}}

\newcommand{\tquatreunn}{
\parbox{6mm}{
\begin{fmfchar*}(6,10)
  \fmfforce{(0.4w,0.1h)}{i1}
  \fmfforce{(0.1w,0.5h)}{o1}
  \fmfforce{(0.5w,0.9h)}{o2}
  \fmfforce{(0.9w,0.9h)}{o3}
  \fmfforce{(0.7w,0.5h)}{v1}
  \fmf{vanilla}{i1,o1}
  \fmf{vanilla}{i1,v1}
  \fmf{vanilla}{v1,o2}
  \fmf{vanilla}{v1,o3}
  \fmfv{decor.shape=circle,decor.filled=1,decor.size=20}{i1}
  \fmfv{decor.shape=circle,decor.filled=1,decor.size=20}{v1}
  \fmfv{decor.shape=circle,decor.filled=0,decor.size=20}{o1}
  \fmfv{decor.shape=circle,decor.filled=0,decor.size=20}{o2}
  \fmfv{decor.shape=circle,decor.filled=1,decor.size=20}{o3}
\end{fmfchar*}
}}

\newcommand{\tquatreunb}{
\parbox{6mm}{
\begin{fmfchar*}(6,10)
  \fmfforce{(0.4w,0.1h)}{i1}
  \fmfforce{(0.1w,0.5h)}{o1}
  \fmfforce{(0.5w,0.9h)}{o2}
  \fmfforce{(0.9w,0.9h)}{o3}
  \fmfforce{(0.7w,0.5h)}{v1}
  \fmf{vanilla}{i1,o1}
  \fmf{vanilla}{i1,v1}
  \fmf{vanilla}{v1,o2}
  \fmf{vanilla}{v1,o3}
  \fmfv{decor.shape=circle,decor.filled=0,decor.size=20}{i1}
  \fmfv{decor.shape=circle,decor.filled=1,decor.size=20}{v1}
  \fmfv{decor.shape=circle,decor.filled=0,decor.size=20}{o1}
  \fmfv{decor.shape=circle,decor.filled=0,decor.size=20}{o2}
  \fmfv{decor.shape=circle,decor.filled=1,decor.size=20}{o3}
\end{fmfchar*}
}}
\newcommand{\tquatreunnn}{
\parbox{6mm}{
\begin{fmfchar*}(6,10)
  \fmfforce{(0.4w,0.1h)}{i1}
  \fmfforce{(0.1w,0.5h)}{o1}
  \fmfforce{(0.5w,0.9h)}{o2}
  \fmfforce{(0.9w,0.9h)}{o3}
  \fmfforce{(0.7w,0.5h)}{v1}
  \fmf{vanilla}{i1,o1}
  \fmf{vanilla}{i1,v1}
  \fmf{vanilla}{v1,o2}
  \fmf{vanilla}{v1,o3}
  \fmfv{decor.shape=circle,decor.filled=1,decor.size=20}{i1}
  \fmfv{decor.shape=circle,decor.filled=1,decor.size=20}{v1}
  \fmfv{decor.shape=circle,decor.filled=1,decor.size=20}{o1}
  \fmfv{decor.shape=circle,decor.filled=1,decor.size=20}{o2}
  \fmfv{decor.shape=circle,decor.filled=1,decor.size=20}{o3}
\end{fmfchar*}
}}

\newcommand{\tquatredeuxn}{
\parbox{6mm}{
\begin{fmfchar*}(6,10)
  \fmfforce{(0.6w,0.1h)}{i1}
  \fmfforce{(0.9w,0.5h)}{o1}
  \fmfforce{(0.5w,0.9h)}{o2}
  \fmfforce{(0.1w,0.9h)}{o3}
  \fmfforce{(0.3w,0.5h)}{v1}
  \fmf{vanilla}{i1,o1}
  \fmf{vanilla}{i1,v1}
  \fmf{vanilla}{v1,o2}
  \fmf{vanilla}{v1,o3}
  \fmfv{decor.shape=circle,decor.filled=1,decor.size=20}{i1}
  \fmfv{decor.shape=circle,decor.filled=0,decor.size=20}{v1}
  \fmfv{decor.shape=circle,decor.filled=1,decor.size=20}{o1}
  \fmfv{decor.shape=circle,decor.filled=1,decor.size=20}{o2}
  \fmfv{decor.shape=circle,decor.filled=0,decor.size=20}{o3}
\end{fmfchar*}
}}

\newcommand{\tquatredeuxb}{
\parbox{6mm}{
\begin{fmfchar*}(6,10)
  \fmfforce{(0.6w,0.1h)}{i1}
  \fmfforce{(0.9w,0.5h)}{o1}
  \fmfforce{(0.5w,0.9h)}{o2}
  \fmfforce{(0.1w,0.9h)}{o3}
  \fmfforce{(0.3w,0.5h)}{v1}
  \fmf{vanilla}{i1,o1}
  \fmf{vanilla}{i1,v1}
  \fmf{vanilla}{v1,o2}
  \fmf{vanilla}{v1,o3}
  \fmfv{decor.shape=circle,decor.filled=0,decor.size=20}{i1}
  \fmfv{decor.shape=circle,decor.filled=0,decor.size=20}{v1}
  \fmfv{decor.shape=circle,decor.filled=1,decor.size=20}{o1}
  \fmfv{decor.shape=circle,decor.filled=1,decor.size=20}{o2}
  \fmfv{decor.shape=circle,decor.filled=0,decor.size=20}{o3}
\end{fmfchar*}
}}

\newcommand{\tquatredeuxnn}{
\parbox{6mm}{
\begin{fmfchar*}(6,10)
  \fmfforce{(0.6w,0.1h)}{i1}
  \fmfforce{(0.9w,0.5h)}{o1}
  \fmfforce{(0.5w,0.9h)}{o2}
  \fmfforce{(0.1w,0.9h)}{o3}
  \fmfforce{(0.3w,0.5h)}{v1}
  \fmf{vanilla}{i1,o1}
  \fmf{vanilla}{i1,v1}
  \fmf{vanilla}{v1,o2}
  \fmf{vanilla}{v1,o3}
  \fmfv{decor.shape=circle,decor.filled=1,decor.size=20}{i1}
  \fmfv{decor.shape=circle,decor.filled=1,decor.size=20}{v1}
  \fmfv{decor.shape=circle,decor.filled=1,decor.size=20}{o1}
  \fmfv{decor.shape=circle,decor.filled=1,decor.size=20}{o2}
  \fmfv{decor.shape=circle,decor.filled=1,decor.size=20}{o3}
\end{fmfchar*}
}}

\newcommand{\tcinqnn}{
\parbox{6mm}{
\begin{fmfchar*}(10,10)
  \fmfforce{(0.5w,0.1h)}{i1}
  \fmfforce{(0.1w,0.9h)}{o1}
  \fmfforce{(0.38w,0.9h)}{o2}
  \fmfforce{(0.62w,0.9h)}{o3}
  \fmfforce{(0.9w,0.9h)}{o4}
  \fmfforce{(0.25w,0.5h)}{v1}
  \fmfforce{(0.75w,0.5h)}{v2}
  \fmf{vanilla}{i1,v1}
  \fmf{vanilla}{i1,v2}
  \fmf{vanilla}{v1,o1}
  \fmf{vanilla}{v1,o2}
  \fmf{vanilla}{v2,o3}
  \fmf{vanilla}{v2,o4}
  \fmfv{decor.shape=circle,decor.filled=1,decor.size=20}{i1}
  \fmfv{decor.shape=circle,decor.filled=1,decor.size=20}{v1}
  \fmfv{decor.shape=circle,decor.filled=1,decor.size=20}{v2}
  \fmfv{decor.shape=circle,decor.filled=1,decor.size=20}{o1}
  \fmfv{decor.shape=circle,decor.filled=1,decor.size=20}{o2}
  \fmfv{decor.shape=circle,decor.filled=1,decor.size=20}{o3}
  \fmfv{decor.shape=circle,decor.filled=1,decor.size=20}{o4}
\end{fmfchar*}
}}

\newcommand{\tcinqn}{
\parbox{6mm}{
\begin{fmfchar*}(10,10)
  \fmfforce{(0.5w,0.1h)}{i1}
  \fmfforce{(0.1w,0.9h)}{o1}
  \fmfforce{(0.38w,0.9h)}{o2}
  \fmfforce{(0.62w,0.9h)}{o3}
  \fmfforce{(0.9w,0.9h)}{o4}
  \fmfforce{(0.25w,0.5h)}{v1}
  \fmfforce{(0.75w,0.5h)}{v2}
  \fmf{vanilla}{i1,v1}
  \fmf{vanilla}{i1,v2}
  \fmf{vanilla}{v1,o1}
  \fmf{vanilla}{v1,o2}
  \fmf{vanilla}{v2,o3}
  \fmf{vanilla}{v2,o4}
  \fmfv{decor.shape=circle,decor.filled=1,decor.size=20}{i1}
  \fmfv{decor.shape=circle,decor.filled=0,decor.size=20}{v1}
  \fmfv{decor.shape=circle,decor.filled=1,decor.size=20}{v2}
  \fmfv{decor.shape=circle,decor.filled=0,decor.size=20}{o1}
  \fmfv{decor.shape=circle,decor.filled=1,decor.size=20}{o2}
  \fmfv{decor.shape=circle,decor.filled=0,decor.size=20}{o3}
  \fmfv{decor.shape=circle,decor.filled=1,decor.size=20}{o4}
\end{fmfchar*}
}}

\newcommand{\tcinqb}{
\parbox{6mm}{
\begin{fmfchar*}(10,10)
  \fmfforce{(0.5w,0.1h)}{i1}
  \fmfforce{(0.1w,0.9h)}{o1}
  \fmfforce{(0.38w,0.9h)}{o2}
  \fmfforce{(0.62w,0.9h)}{o3}
  \fmfforce{(0.9w,0.9h)}{o4}
  \fmfforce{(0.25w,0.5h)}{v1}
  \fmfforce{(0.75w,0.5h)}{v2}
  \fmf{vanilla}{i1,v1}
  \fmf{vanilla}{i1,v2}
  \fmf{vanilla}{v1,o1}
  \fmf{vanilla}{v1,o2}
  \fmf{vanilla}{v2,o3}
  \fmf{vanilla}{v2,o4}
  \fmfv{decor.shape=circle,decor.filled=0,decor.size=20}{i1}
  \fmfv{decor.shape=circle,decor.filled=0,decor.size=20}{v1}
  \fmfv{decor.shape=circle,decor.filled=1,decor.size=20}{v2}
  \fmfv{decor.shape=circle,decor.filled=0,decor.size=20}{o1}
  \fmfv{decor.shape=circle,decor.filled=1,decor.size=20}{o2}
  \fmfv{decor.shape=circle,decor.filled=0,decor.size=20}{o3}
  \fmfv{decor.shape=circle,decor.filled=1,decor.size=20}{o4}
\end{fmfchar*}
}}

\section{Introduction}

Renormalization theory has been recently revitalized by the discovery
a Hopf algebra that transforms the dreadful combinatorics of 
renormalization into
a mechanical application of the Hopf algebra properties of rooted trees
\cite{Kreimer98,Kreimer,Connes}. 

In the companion paper \cite{Brouder}, Butcher's theory has been
presented as an alternative way to describe the Hopf structure of 
the algebra of renormalization. A particularly useful aspect of 
Butcher's approach is that solutions of non-linear differential
equations can be written as a sum over rooted trees.

In the present paper, Butcher's strategy is adapted to 
equations involving functional derivatives, that were first proposed
by Schwinger \cite{Schwinger} and will be called Schwinger
equations in the rest of the paper. 
Schwinger equations are not commonly considered as a useful tool for
computation. The purpose of this article is to show that,
by using series over planary binary trees, Schwinger equations
can be turned into explicit calculation methods.

The series we manipulate are indexed by planar binary
trees. So we first present
some basic properties of planar binary trees.
Then the solution of simple Schwinger equations
are written as a sum over planar binary trees, with
recursively defined coefficients. 
To make a comparison with power series, the Schwinger equation
would correspond to a differential equation for the sum
of the series, whereas the formula we put forward corresponds
to a recursive definition of the terms of the series: it 
does not contain so much information as the Schwinger equation,
but it is more explicit to calculate the terms of the
series.

As an example, the full fermion
and photon propagators of quantum electrodynamics (QED)
are written as a sum over planar binary trees.
Renormalization is briefly described and
applications are given for QED with an external source,
the vacuum polarization, the fermion self-energy and the
irreducible vertex.

\section{Planar binary trees}
In contrast to Ref.\cite{Brouder}, we do not use rooted trees 
but planar 
binary trees. Both can be drawn on a plane, but no permutation 
of vertices is
allowed for planar trees. 

As an example, $\tquatreunnn$ and $\tquatredeuxnn$ are two different
planar trees, although they represent the same rooted tree.

In planar trees, we distinguish two types of vertices: 
the leaves (which have no sons) and the remaining vertices 
(including the root), which we call internal vertices.
In planar binary trees, internal 
vertices have exactly two sons.

We now follow the notation of Loday and Ronco \cite{LodayRonco}.
Planar binary trees have an odd number of vertices. We denote 
$Y_n$ the set of planar binary trees with $2n+1$ vertices. If
$t$ belongs to $Y_n$, $t$ has $n+1$ leaves and $n$ internal vertices.
The number of elements of $Y_n$ is $\frac{(2n)!}{n!(n+1)!}$:
the Catalan numbers, which enter many combinatorial problems
\cite{Conway} and should probably be called Ming numbers
\cite{Larcombe}.
If $t_1\in Y_m$ and $t_2\in Y_n$, the grafting of $t_1$ and $t_2$
is the tree $t\in Y_{m+n+1}$ obtained by putting $t_1$ on the left
of $t_2$ and by joining the roots of $t_1$ and $t_2$ to a new vertex
that becomes the root of $t$. This operation is denoted
by $t=t_1\vee t_2$.
For instance, grafting $\tunn$ and $\ttroisnn$ gives 
$\tunn\vee\ttroisnn=\tquatreunnn$.

In the companion paper, rooted trees were graded by
the number of their vertices. Here, planar binary trees have
an odd number of vertices, and it is more natural to
grade them differently: for each tree $t$, we
define $|t|$ as the integer $n$ such that $t\in Y_n$.
Thus, a tree $t$ has $2|t|+1$ vertices.

An essential property of planar binary trees is that each tree $t$
different from $\tunn$ can be written in a unique way as $t_1\vee t_2$,
where $t_1$ and $t_2$ are called the branches of $t$.
Moreover, grafting provides a recursive definition of 
planar binary trees \cite{Loday}:
\begin{eqnarray}
Y_n &=&  \overset{n-1}{\underset{k=0}{\bigcup}}
 Y_k \vee Y_{n-k-1},\quad Y_0=\{\tunn\}. \label{dectree}
\end{eqnarray}
The notation $Y_k \vee Y_{n-k-1}$ means that all the trees of $Y_k$
are grafted with all the trees of $Y_{n-k-1}$.

Planar binary trees have received much attention recently because of their
relation to new algebraic structures \cite{Loday,Frabetti}.

\section{Schwinger equations}

In this section, the solution of a linear Schwinger
equation is given as a sum over planar binary trees.
But we first introduce the concept of a functional
derivative.

A functional $A(\phi)$ is defined loosely as a map sending a
distribution $\phi$ to a complex number (see Ref.\cite{Glimm}
for details). If $\psi$ is a distribution, the 
functional derivative of $A(\phi)$ in the 
direction $\psi$ is defined 
as the limit for $\epsilon\rightarrow 0$ of 
$(A(\phi+\epsilon\psi)-A(\phi))/\epsilon$.
Finally, the functional derivative of $A(\phi)$ 
with respect to $\phi(x)$,
$\frac{\delta A(\phi)}{\delta\phi(x)}$,
is defined as the functional derivative of 
$A(\phi)$ in the direction 
$\delta_x$, where $\delta_x$ is the Dirac function 
$\delta_x(y)=\delta(y-x)$.

\subsection{Examples of functional derivatives\label{funcdersection}}
A classical example is $A(\phi)=\int dx f(x)\phi(x)$, giving
easily
$\frac{\delta A(\phi)}{\delta \phi(x)}= f(x)$.
A further example, that will be useful in the sequel, is
$A(\phi)=\int dxdy f(x,y)\phi(x)\phi(y)$. Then
\begin{eqnarray*}
\frac{\delta A(\phi)}{\delta \phi(x)} &=& \int dy f(x,y)\phi(y)+
  \int dy f(y,x)\phi(y),\\
\frac{\delta^2 A(\phi)}{\delta \phi(x)\delta \phi(y)} &=& 
\frac{\delta}{\delta \phi(y)}
\frac{\delta A(\phi)}{\delta \phi(x)}=
f(x,y)+f(y,x).
\end{eqnarray*}
More generally, if 
\begin{eqnarray*}
A(\phi)&=& \int dx_1 \dots dx_n f(x_1,\dots,x_n) \phi(x_1)\dots \phi(x_n),\\
\end{eqnarray*}
then
\begin{eqnarray*}
\frac{\delta^n A(\phi)}{\delta \phi(x_1)\cdots \delta \phi(x_n)} &=& 
\sum_{\sigma\in\mathcal{S}_n} f(x_{\sigma(1)},\dots,x_{\sigma(n)}),
\end{eqnarray*}
where $\mathcal{S}_n$ is the set of permutations of $n$ elements.

In practice, $A(\phi)$ is often a Green function.
Take the Green function defined by
$(\Delta_x - \phi(x))A(\phi;x,y)=\delta(x-y)$, which
can be written $A(\phi)=(\Delta-\phi)^{-1}$.
To calculate the functional derivative, we put
$A(\phi+\epsilon\psi)=(\Delta-\phi-\epsilon\psi)^{-1}$.
The operator identity $Y^{-1}=X^{-1}+X^{-1}(X-Y)Y^{-1}$
gives us
$A(\phi+\epsilon\psi)=A(\phi)+\epsilon A(\phi)\psi A(\phi+\epsilon\psi)$.
Thus, taking the limit $\epsilon\rightarrow 0$,
\begin{eqnarray*}
\frac{\delta A(\phi;x,y)}{\delta \psi}=
\int ds A(\phi;x,s)\psi(s)A(\phi;s,y).
\end{eqnarray*}
If we choose now the distribution $\psi(s)=\delta(s-z)$ we find
\begin{eqnarray}
\frac{\delta A(\phi;x,y)}{\delta \phi(z)}= A(\phi;x,z)A(\phi;z,y).
\label{deltaGreen}
\end{eqnarray}
This identity will be used repeatedly in the sequel.

\subsection{A simple Schwinger equation}
As an introduction to the method of planar binary trees we consider the
Schwinger equation
\begin{eqnarray}
X=A+F(X,\frac{\delta X}{\delta v(z)}), \label{lineq}
\end{eqnarray}
where $A$ is a functional of $v$, $F$ is linear
in $X$ and $\frac{\delta X}{\delta v(z)}$, and $z$ is a variable over which
$F$ integrates. In an equation like (\ref{lineq}), $F$ is called the 
integral operator of the equation and $A$ is called its initial data.

A common example of such an equation is obtained when $X=X(x,y)$,
the initial data $A(x,y)$ are the Green function 
$(\Delta_x-v(x))A(x,y)=\delta(x-y)$ discussed in section 
\ref{funcdersection}
and the integral operator is 
\begin{eqnarray*}
F(X,\frac{\delta X}{\delta v(z)})(x,y) &=& \int ds dz X(x,s) f(s,z) 
  \frac{\delta X(s,y)}{\delta v(z)},
\end{eqnarray*}
for some function $f(s,z)$.

Such a Schwinger equation summarizes an infinity of equations
that can be obtained by taking successive functional
derivatives of Eq.(\ref{lineq}) with respect to $v(z)$.
\begin{eqnarray*}
X&=&A+F(X,\frac{\delta X}{\delta v(z)}),\\
\frac{\delta X}{\delta v(z_1)}&=&\frac{\delta A}{\delta v(z_1)}+
  F(\frac{\delta X}{\delta v(z_1)},\frac{\delta X}{\delta v(z)})+
  \\&&\hspace*{5mm}
  F(X,\frac{\delta^2 X}{\delta v(z_1)\delta v(z)}),\\
& \cdots &
\end{eqnarray*}
When we take the $n$-th functional derivative of both sides of the
($n$-$1$)-th equation with respect to $v(z_n)$, the equation gets 
an additional variable $z_n$, and the chain rule is used
to apply $\frac{\delta}{\delta v(z_n)}$ to
the right-hand side 
of the ($n$-$1$)-th equation.

If this is iterated to all values of $n$,
we obtain an infinite system of nonlinear integral equations.
This system seems difficult to solve because the $n$-th differential of $X$
depends on the $k$-th differentials of $X$ for $k=0$ to $n+1$.

\subsection{The series solution \label{solsection}}
To write the solution of Eq.(\ref{lineq}), we must introduce some
notation. Sets of arguments will be often needed, so we write
 $\{z\}_{i,j}=z_i,z_{i+1},\dots,z_j$,
($\{z\}_{i,j}=\varnothing$ if $j<i$). Furthermore, 
if $f(\{z\}_{1,n})=f(z_1,\dots,z_n)$ is a function of $n$
variables, then $f_\Sigma(\{z\}_{1,n})$
is defined as the sum of $n$ terms,
where the first variable $z_1$ is shifted step by step from the 
first to the $n$-th position:
\begin{eqnarray*}
f_\Sigma(\{z\}_{1,n}) &=&
f(\{z\}_{1,n}) + f(z_2,z_1,\{z\}_{3,n})+\cdots+\\&&\hspace*{3mm}
f(\{z\}_{2,n-1},z_1,z_n)+ f(\{z\}_{2,n},z_1).
\end{eqnarray*}

For each planar binary tree $t$, we define an infinite-dimensional
vector $\phi(t)$, 
with components $\phi^n(t)$, where $n$ goes from 0 to infinity.
The $n$-th component is a function of $n$ variables $z_1,\dots,z_n$.
Now we define the initial data.
For $t=\tunn$, we take $\phi^0(\tunn) = A$ and
\begin{eqnarray}
\phi^1(\tunn;z_1) &=& \frac{\delta A}{\delta v(z_1)}.
\label{initial}
\end{eqnarray}
The most natural choice is to define $\phi^n(\tunn)$
as $1/n!$ times the $n$-th
functional derivative of $A$ with respect to $v(z)$, but 
this is not always the most economical
choice in practice. The only condition that we need 
for $n>1$ is
\begin{eqnarray}
\frac{\delta \phi^{n-1}(\tunn;\{z\}_{1,n-1})}{\delta v(z_n)} &=&
 \phi^{n}_\Sigma(\tunn;\{z\}_{1,n}).
\label{condition}
\end{eqnarray}
When $A(x,y)$ is a Green function of the kind discussed in 
section \ref{funcdersection},
it is not difficult to build such a $\phi(\tunn)$ from given
initial data $A$. We can use Eq.(\ref{deltaGreen}) to show that
\begin{eqnarray}
\phi^0(\tunn) &=& A(x,y),\nonumber\\
\phi^1(\tunn;z_1) &=& A(x,z_1)A(z_1,y),\nonumber\\
&\dots&\nonumber\\
\phi^{n}(\tunn;\{z\}_{1,n})&=&
  A(x,z_1)A(z_1,z_2)\dots A(z_n,y),\label{AAA}
\end{eqnarray}
satisfies Eq.(\ref{initial}) and condition (\ref{condition}).

With this notation we can now write the solution of Eq.(\ref{lineq}) as
\begin{eqnarray}
X&=& \sum_{t} \phi^0 (t), \label{Xsumphi}
\end{eqnarray}
where $t$ spans the set of planar binary trees.
Moreover
\begin{eqnarray*}
\frac{\delta X}{\delta v(z_1)}&=&
\sum_{t} \phi^1 (t;z_1),\\
&\dots&\\
\frac{\delta^n X}{\delta v(z_1)\cdots\delta v(z_n)}&=&
\sum_{\sigma\in\mathcal{S}_n}
\sum_{t} \phi^n (t;z_{\sigma(1)},\dots,z_{\sigma(n)}).
\end{eqnarray*}

For each planar binary tree $t$, the vector $\phi(t)$ is 
calculated
as a function of the vectors $\phi(t_1)$ and $\phi(t_2)$,
where $t_1$ and $t_2$ are the branches of $t$. Since
$\phi(\tunn)$ is defined from Eq.(\ref{initial}),
this defines $\phi(t)$ recursively.
The recursive definition of $\phi(t)$ is given explicitly by
\begin{eqnarray}
\phi^n(t;\{z\}_{1,n}) &=& \sum_{k=0}^n
  F\big(\phi^k(t_1;\{z\}_{1,k}),\nonumber\\&&\hspace*{10mm}
  \phi^{n-k+1}_\Sigma(t_2;z,\{z\}_{k+1,n})\big), \label{sumlineq}
\end{eqnarray}
for $t\not=\tunn$
and $\phi^n(\tunn)$ is defined in Eq.(\ref{AAA}).

In a quantum field interpretation, $\tunn$ represents the bare fields,
$\ttroisnn$ represents the interaction, and the sum over trees represents
all the combinations of interaction that gives the full propagator.

A proof of Eqs.(\ref{Xsumphi}) and (\ref{sumlineq}) is given in the appendix.

\subsection{Enumeration \label{enumsection}}
If the initial data $A$ is such that $\phi^n(\tunn)$ has only one term,
as in Eq.(\ref{AAA}),
the chain rule applied to the functional derivative gives a number
of terms for $\phi^n(t)$ that we denote $|\phi^n(t)|$.
Equation (\ref{sumlineq}) gives us the following recurrence relation
for  $|\phi^n(t)|$:
\begin{eqnarray*}
|\phi^n(t)|&=&\sum_{k=0}^n (n-k+1) |\phi^k(t_1)| |\phi^{n-k+1}(t_2)|,\\
|\phi^n(\tunn)|&=&1.
\end{eqnarray*}
Using the binomial identity
\begin{eqnarray*}
\sum_{k=0}^n {a+k \choose a}{b+n-k \choose b}&=&{a+b+n+1 \choose a+b+1},
\end{eqnarray*}
it can be shown that the solution of this equation is 

\begin{eqnarray*}
|\phi^n(t)|&=& \bar{\varphi}(t) {2|t|+n \choose 2|t|},
\end{eqnarray*}
where $\bar{\varphi}(t)$ is an integer which depends only on the tree $t$ 
(not on $n$) and is defined recursively by
\begin{eqnarray*}
\bar{\varphi}(t)&=& \bar{\varphi}(t_1)(2|t_2|+1)\bar{\varphi}(t_2),\\
\bar{\varphi}(\tunn)&=& 1.
\end{eqnarray*}
$t_1$ and $t_2$ are the two branches of $t$.

\subsection{A compact notation \label{compactsection}}
To write a compact expression for the recursive definition
of $\phi(t)$ we define the deconcatenation of 
$(z_1,\dots,z_n)$ by \cite{Loday99}
\begin{eqnarray*}
\Delta(z_1,\dots,z_n)=\sum_{i=0}^{n} (z_1,\dots,z_i)
\otimes (z_{i+1},\dots,z_n).
\end{eqnarray*}
If $z$ belongs to the vector space $V$, 
the map $\phi(t)$ acts on the tensor module (Fock space)
$T(V)= \overset{\infty}{\underset{n=0}{\bigoplus}}
  V^{\otimes n}$.
Sometimes, as in QED, $\phi$ is defined on
$T(V)\times M$, where $M$ is a fixed vector space,
for instance $M=V^2$ for the photon propagator.

We define the operator $d(z)$ by
\begin{eqnarray*}
d(z) \phi^n(t;z_1,\dots,z_n) = \phi^{n+1}_\Sigma(t;z,z_1,\dots,z_n).
\end{eqnarray*}

The recursive definition of $\phi$ becomes
\begin{eqnarray*}
\phi(t)=F\circ\big(Id\otimes d(z))\circ
  (\phi(t_1)\otimes\phi(t_2))\circ\Delta,
\end{eqnarray*}
where $F(a\otimes b)=F(a,b)$.

It would be interesting to find a family of equations
such that, for any $\phi$ from $T(V)$ to $\mathbb{C}$,
there is a member of the family of which $\phi$ is a
solution. This would generalize Butcher's density
theorem\cite{Butcher}, and would provide a general 
class of equations
that would be satisfied by the renormalized 
Green functions.

\subsection{Algebra structure \label{groupsection}}
In the case of rooted trees, Butcher \cite{Butcher72} has 
defined a group structure of 
Runge-Kutta methods that Hairer and 
Wanner \cite{Hairer74} interpreted as
a composition of Butcher series. A similar approach can 
be used for planar binary trees.
A powerful aspect of Butcher's approach is that algebraic
operations are defined on two spaces at the same time:
the space of Runge-Kutta methods, and the space of
maps over trees.
The same strategy will be used here, and the operations
will be defined on the space of integral operators and
on the space of maps over planar binary trees.

We start with the addition.
If we have two Schwinger equations 
$X=A+F(X,\frac{\delta X}{\delta v})$ and 
$Y=B+G(Y,\frac{\delta Y}{\delta v})$, the addition of the
integral operators is $H=F+G$
and the addition of the maps $\phi$ (corresponding to the
first equation) and $\psi$ (for the second equation) is
defined recursively by
$\chi^n(\tunn) = \phi^n(\tunn)+\psi^n(\tunn)$, for
the initial data $A+B$ and
\begin{eqnarray*}
\chi^n(t;\{z\}_{1,n}) &=& \\&&\hspace*{-20mm}\sum_{k=0}^n
  F\big(\chi^k(t_1;\{z\}_{1,k}),
  \chi^{n-k+1}_\Sigma(t_2;z,\{z\}_{k+1,n})\big) + 
   \\&&\hspace*{-12mm}\sum_{k=0}^n
  G\big(\chi^k(t_1;\{z\}_{1,k}),
  \chi^{n-k+1}_\Sigma(t_2;z,\{z\}_{k+1,n})\big).
\end{eqnarray*}
This addition defines clearly a commutative group structure for
the integral operators. It defines also a commutative group
structure for the space of maps, where the unit element
is $\phi(t)=0$ for all $t$, and the opposite of
$\phi(t)$ is $\psi(t)=-(-1)^{|t|}\phi(t)$.

Multiplication by a scalar $\lambda$ is defined similarly.
An integral operator $F$ becomes $\lambda F$, and the
corresponding map $\phi(t)$ becomes $\lambda^{|t|}\phi(t)$.
Notice that maps are not equivalent to integral operators
since they contain also the initial data. The present
definition of the multiplication by a scalar corresponds
to the case where the initial data are not changed.
If the initial data are also multiplied by $\lambda$,
then $\phi(t)$ becomes $\lambda^{|t|+1}\phi(t)$.

This addition is useful when a Schwinger equation is the
sum of various terms. Addition and multiplication by a
scalar is probably all we need to renormalize quantum
electrodynamics. However, we can proceed with Butcher's
strategy and define another operation coming from 
a composition of solutions.
If we start from two Schwinger equations 
$X=A+F(X,\frac{\delta X}{\delta v})$ and 
$Y=B+G(Y,\frac{\delta Y}{\delta v})$, 
the composition of the 
solutions is defined as the $Y$ obtained with the 
initial data $B=X$. 

It is shown in the appendix that, if $\chi$ is the map corresponding
to $Y$ (i.e. $Y=\sum_t \chi(t)$), then $\chi(t)=\phi(t)+\psi(t)$,
where $\phi(t)$ is the map associated to the equation for $X$
and $\psi^n(t)$ is given by $\psi^n(\tunn) = 0$ and
\begin{eqnarray}
\psi^n(t;\{z\}_{1,n}) &=& \sum_{k=0}^n 
  G\big(\phi^k(t_1;\{z\}_{1,k})+\psi^k(t_1;\{z\}_{1,k}),
\nonumber\\&&\hspace*{-22mm}
  \phi_\Sigma^{n-k+1}(t_2;z;\{z\}_{k+1,n})+
  \psi_\Sigma^{n-k+1}(t_2;z;\{z\}_{k+1,n})\big).\label{compeq}
\end{eqnarray}
This defines a product of integral operators and of maps.
In the first proof of Eq.(\ref{compeq}) given in the appendix,
the integral operator corresponding to this product is
constructed. Notice that this operator acts on vectors
${X\choose X+Y}$. This product has a unit element (given by
$G=0$). 

In the next section, the present method will be applied to
the example of quantum electrodynamics (QED).

\section{The case of QED}

We work in the flat Minkowski space with a diagonal
metric $g$ (the diagonal is $(1,-1,-1,-1)$).
The electron charge is $e=-|e|$. Repeated indices are summed over.

In 1951, Schwinger \cite{Schwinger} devised coupled equations
involving functional derivatives of $S(x,y;J)$,
the full fermion
propagator of QED in the presence of an external 
electromagnetic source $J_\mu(x)$:
\begin{eqnarray*}
[ {\Box} g_{\mu\nu} -(1-\xi) \partial_{\mu} 
\partial_{\nu} ] A^{\nu}(x;J) =
-J_\mu(x)-\\\hspace*{20mm}
  i e \mathrm{tr}[\gamma_\mu S(x,x;J)],
\end{eqnarray*}
\begin{eqnarray}
\Big[i\gamma^\mu\partial_\mu-m-e\gamma^\mu A_\mu(x;J)
 +ie\gamma^\mu \frac{\delta}{\delta J_\mu(x)} \Big] S(x,y;J) 
   &=&\nonumber\\\hspace*{-22mm} \delta(x-y). \label{Schwinger}
\end{eqnarray}

Building on a work by Polivanov \cite{Polivanov}, Bogoliubov
and Shirkov \cite{Bogoliubov} transformed 
this equation into a Schwinger equation coupling the full fermion
propagator $S(x,y)$ with the full photon propagator $D_{\mu\nu}(x,y)$:
\begin{eqnarray}
[\Box g^{\mu\nu} -(1-\xi) \partial^\mu\partial^\nu] D_{\nu\rho}(x,y)
&=& g^\mu_\rho \delta(x-y)-
\nonumber\\&& \hspace*{-45mm}
 i e \int d^4z\,\mathrm{tr}\big[\gamma^\mu  
  \frac{\delta S(x,x;A)}{\delta A_{\nu}(z)}
 \big] D_{\nu\rho}(z,y;A),\label{IZ1}\\ 
\big[i\gamma^\mu\partial_\mu-m-e\gamma^\mu A_\mu(x)\big] S(x,y;A)
 &=&\delta(x-y)+
\nonumber\\&& \hspace*{-45mm}
 ie\int d^4z\,\gamma^\mu   
  D_{\mu\rho}(x,z;A)\frac{\delta S(x,y;A)}{\delta
A_{\rho}(z)}, \label{IZ2}
\end{eqnarray}
where $A(x)$ is now an external electromagnetic field. As explained
in Ref.\cite{Bogoliubov},
Eqs.(\ref{IZ1}) and (\ref{IZ2}) are not completely equivalent to 
Eq.(\ref{Schwinger}),
they are valid in the limit $A=0$ ($J=0$), 
which is the standard case of QED.

Multiplying Eq.(\ref{IZ1}) by the bare photon propagator, $D^0_{\mu\nu}(x,y)$
and Eq.(\ref{IZ2}) by 
the bare fermion propagator in the presence of $A$,
$S^0(x,y;A)=[i\gamma^\mu\partial_\mu-m-e\gamma^\mu A_\mu]^{-1}$,
we obtain our starting Schwinger equations:
\begin{eqnarray}
D_{\mu\nu}(x,y;A) &=& D^0_{\mu\nu}(x,y)-
          ie\int d^4z d^4z' D^0_{\mu\lambda}(x,z)\nonumber\\&&
  \hspace*{6mm}\mathrm{tr}\big[\gamma^\lambda 
  \frac{\delta S(z,z;A)}{\delta A_{\lambda'}(z')} 
   \big] D_{\lambda'\nu}(z',y;A),\nonumber\\
S(x,y;A) &=& S^0(x,y;A)+ie\int d^4z d^4z' S^0(x,z;A)\nonumber\\&&
  \hspace*{6mm}
  \gamma^\lambda D_{\lambda\lambda'}(z,z';A)
  \frac{\delta S(z,y;A)}{\delta A_{\lambda'}(z')}.
\label{eqbase}
\end{eqnarray}
In principle, these equations fully determine 
$S(x,y;A)$ and $D_{\mu\nu}(x,y;A)$.

\subsection{The tree solution \label{treesolsection}}

The method of the previous section is now used to write the solution
of Eq.(\ref{eqbase}). Since $S^0(x,z;A)$
depends on the external potential $A(x)$, the small extension presented
in the appendix is required. All quantities will be taken
at $A(x)=0$, so the external potential will not be mentioned
(e.g. $S(x,y)$ means $S(x,y;A)$ at $A=0$).

The notation
$\{\lambda, z\}_{i,j}=\lambda_i,z_i,\lambda_{i+1},z_{i+1},\dots,\lambda_j,z_j$
enables us to write the  solution as
\begin{eqnarray}
S(x,y)&=& \sum_{t} e^{2|t|} \phi^0 (t^\bullet;x,y),\label{fermiontreesol}\\
\frac{\delta S(x,y)}{\delta A_{\lambda_1}(z_1)}&=&
\sum_{t} e^{2|t|+1} 
  \phi^1 (t^\bullet;x,y;\lambda_1 z_1),\nonumber\\
D_{\mu\nu}(x,y)&=&
\sum_{t} e^{2|t|} \phi^0_{\mu\nu} (t^\circ;x,y),\label{photontreesol}\\
\frac{\delta^n S(x,y)}{\delta A_{\lambda_1}(z_1)\cdots
   \delta A_{\lambda_n}(z_n)}&=&
\sum_{\sigma\in \mathcal{S}_n}\sum_{t} e^{2|t|+n} \nonumber\\&&
  \hspace*{5mm}
  \phi^n (t^\bullet;x,y;\{\lambda,z\}_{\sigma(1),\sigma(n)}),\nonumber\\
\frac{\delta^n D_{\mu\nu}(x,y)}{\delta A_{\lambda_1}(z_1)
   \cdots\delta A_{\lambda_n}(z_n)}&=&
 \sum_{\sigma\in \mathcal{S}_n} \sum_{t} e^{2|t|+n} \nonumber\\&&
  \hspace*{5mm}
   \phi^n_{\mu\nu} (t^\circ;x,y;\{\lambda,z\}_{\sigma(1),\sigma(n)}).\nonumber
\end{eqnarray}
Another change in the notation is that trees have now two colors. 
The map $\phi(t)$ has a supersymmetric flavor because it has a 
fermion component
for the full fermion propagator and a photon component for the
full photon propagator. 
It is convenient to transfer the fermion/photon index from 
$\phi$ to the tree $t$.
Thus, $\phi(t^\circ)$ is the photon component of $\phi(t)$ and
$\phi(t^\bullet)$ is its fermion component.
Futhermore, $\phi(t^\circ)$ will be drawn with a white root and
$\phi(t^\bullet)$ with a black root. 
All trees are built as $t_1^\circ \vee t_2^\bullet$, where the white (photon) branch
is on the left and the black (fermion) branch is on the right.

For example, the fermion trees are
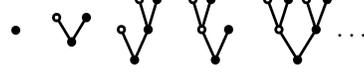
\begin{eqnarray*}
\tunn\quad\ttroisn\quad\tquatreunn\quad\tquatredeuxn\quad\tcinqn\quad\dots
\end{eqnarray*}
and the photon trees are
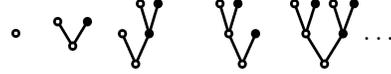
\begin{eqnarray*}
\tunb\quad\ttroisb\quad\tquatreunb\qquad\tquatredeuxb\quad\tcinqb\quad\dots
\end{eqnarray*}

Frabetti has calculated the number of trees with $2p+2q+1$ vertices, 
$p+1$ white leaves
and $q+1$ black leaves as \cite{Frabetti}
\begin{eqnarray*}
c_{p,q}=\frac{(p+q)!}{p!q!}\frac{(p+q+1)!}{(p+1)!(q+1)!}.
\end{eqnarray*}

The following recursive definition 
gives $\phi(t)$ in terms of $\phi(t_1^\circ)$ and $\phi(t_2^\bullet)$, where
$t_1^\circ$ and $t_2^\bullet$ are the branches of $t$:
\begin{eqnarray}
\phi^n(t^\bullet;x,y;\{\lambda, z\}_{1,n})&=& 
 i\sum_{k=0}^{n}\sum_{k'=0}^{n-k}\int d^4z d^4z'\nonumber\\&&\hspace*{-2cm}
 \phi^{n-k-k'}(\tunn;x,z;\{\lambda, z\}_{1,n-k-k'})\nonumber\\&&\hspace*{-2cm}
 \gamma^\lambda
  \phi^k_{\lambda\lambda'} (t_1^\circ;z,z';\{\lambda,z\}_{n-k-k'+1,n-k'})
  \nonumber\\&&\hspace*{-2cm}
 \phi^{k'+1}_\Sigma(t_2^\bullet;z,y;\lambda',z',\{\lambda,z\}_{n-k'+1,n}),
     \label{phinoir}\\
\phi^n_{\mu\nu}(t^\circ;x,y;\{\lambda, z\}_{1,n})&=& 
 -i\sum_{k=0}^{n}\int d^4z d^4z' D^0_{\mu\lambda}(x,z)\nonumber\\&&
  \hspace*{-25mm}
 \phi^{k}_{\lambda'\nu}(t_1^\circ;z',y;\{\lambda, z\}_{1,k})\nonumber\\&&
  \hspace*{-25mm}
  \mathrm{tr}\big[\gamma^\lambda
  \phi^{n-k+1}_\Sigma(t_2^\bullet;z,z;\lambda' z',\{\lambda, z\}_{k+1,n})
  \big]. \label{phiblanc}
\end{eqnarray}

This recursive definition is completed
by giving the components of $\phi^n(\tunn)$
and $\phi^n(\tunb)$:
\begin{eqnarray*}
\phi^0_{\mu\nu}(\tunb;x,y) &=& D^0_{\mu\nu}(x,y),\\
\phi^n_{\mu\nu}(\tunb;x,y) &=& 0\quad\mathrm{for}\quad n\ge 1,\\
\phi^0(\tunn;x,y) &=& S^0(x,y),\\
\phi^1(\tunn;x,y;\lambda_1 z_1) &=& S^0(x,z_1)\gamma^{\lambda_1}S^0(z_1,y),\\
\phi^n(\tunn;x,y;\{\lambda,z\}_{1,n}) &=& S^0(x,z_1)\gamma^{\lambda_1}
  S^0(z_1,z_2)\gamma^{\lambda_2}\cdots\\&&\hspace*{6mm}
  \gamma^{\lambda_n}S^0(z_n,y).
\end{eqnarray*}

In practice, the double sum of Eq.(\ref{phinoir}) is replaced by:
\begin{eqnarray}
\phi^n(t^\bullet;x,y;\{\lambda,z\}_{1,n})&=& \nonumber\\&&\hspace*{-3cm}
 S^0(x,z_1)\gamma^{\lambda_1} \nonumber
 \phi^{n-1}(t^\bullet;z_1,y;\{\lambda,z\}_{2,n})+\nonumber\\&&\hspace*{-3cm}
 i\sum_{k=0}^{n}\int d^4z d^4z'
 S^0(x,z) \gamma^\lambda
 \phi^k_{\lambda\lambda'} (t_1^\circ;z,z';\{\lambda,z\}_{1,k})\nonumber\\&&
  \hspace*{-10mm}
 \phi^{n-k+1}_\Sigma(t_2^\bullet;z,y;\lambda' z',\{\lambda,z\}_{k+1,n})
 \label{phinoir2}.
\end{eqnarray}
This is still a recursive definition, but it uses now the smaller components
of $\phi$ for the same tree $t^\bullet$. Notice that the first term of
Eq.(\ref{phinoir2}) is absent if $n=0$.

Three remarks can be useful at this point. Firstly, considering all
quantities at $A=0$, we find as in section 
\ref{funcdersection} that
$\delta S^0(x,y)/\delta A_\lambda(z)=e S^0(x,z)\gamma^\lambda S^0(z,y)$.
Therefore, in the definition of
$\phi^n(\tunn)$, a factor $e^{n}$
was suppressed and transferred to the solution Eqs.(\ref{fermiontreesol}) and
(\ref{photontreesol}). It must be checked that this is compatible
with renormalization.
Secondly, it will be shown in section \ref{Furrysection} that, 
from Furry's theorem,
the components $\phi^n(t^\circ)$ are zero when $n$ is odd. This reduces the
sums in Eqs.(\ref{phinoir}), (\ref{phiblanc}) and (\ref{phinoir2}) to the even
components of $\phi(t_1^\circ)$.
Finally, the vector space $V$ of section \ref{compactsection}
has now become the space $\{0,1,2,3\}\times \mathbb{R}^4$.

\subsection{Diagrammatic interpretation}
The recursive solution of the previous sections can be illustrated in the usual
diagrammatic language:
\begin{eqnarray*}
\phi(\tunb) &=& \left( \begin{array}{c} 
   \parbox{6mm}{
    \begin{fmfchar*}(6,3)
    \fmfforce{(0.1w,0.5h)}{i1}
    \fmfforce{(0.9w,0.5h)}{o1}
    \fmf{photon}{i1,o1}
    \fmfv{decor.shape=circle,decor.filled=1,decor.size=10}{i1}
    \fmfv{decor.shape=circle,decor.filled=1,decor.size=10}{o1}
    \end{fmfchar*}}
    \\ 0 \\ 0 \\ 0 \\ \vdots \end{array}\right)\qquad
\phi(\tunn) = \left( \begin{array}{c} 
   \parbox{6mm}{
    \begin{fmfchar*}(6,3)
    \fmfforce{(0.1w,0.3h)}{i1}
    \fmfforce{(0.9w,0.3h)}{o1}
    \fmf{fermion}{o1,i1}
    \fmfv{decor.shape=circle,decor.filled=1,decor.size=10}{i1}
    \fmfv{decor.shape=circle,decor.filled=1,decor.size=10}{o1}
    \end{fmfchar*}}
    \\ 
   \parbox{12mm}{
    \begin{fmfchar*}(12,8)
    \fmfforce{(0.1w,0.1h)}{i1}
    \fmfforce{(0.9w,0.1h)}{o1}
    \fmfforce{(0.5w,0.8h)}{o2}
    \fmfforce{(0.5w,0.1h)}{v1}
    \fmf{fermion}{o1,v1,i1}
    \fmf{photon}{v1,o2}
    \fmfv{decor.shape=circle,decor.filled=1,decor.size=10}{i1}
    \fmfv{decor.shape=circle,decor.filled=1,decor.size=10}{o1}
    \fmfv{decor.shape=circle,decor.filled=1,decor.size=10}{v1}
    \end{fmfchar*}}
 \\ 
   \parbox{18mm}{
    \begin{fmfchar*}(18,8)
    \fmfforce{(0.1w,0.1h)}{i1}
    \fmfforce{(0.9w,0.1h)}{o1}
    \fmfforce{(0.37w,0.8h)}{o2}
    \fmfforce{(0.63w,0.8h)}{o3}
    \fmfforce{(0.37w,0.1h)}{v1}
    \fmfforce{(0.63w,0.1h)}{v2}
    \fmf{fermion}{o1,v2,v1,i1}
    \fmf{photon}{v1,o2}
    \fmf{photon}{v2,o3}
    \fmfv{decor.shape=circle,decor.filled=1,decor.size=10}{i1}
    \fmfv{decor.shape=circle,decor.filled=1,decor.size=10}{o1}
    \fmfv{decor.shape=circle,decor.filled=1,decor.size=10}{v1}
    \fmfv{decor.shape=circle,decor.filled=1,decor.size=10}{v2}
    \end{fmfchar*}}
 \\ 
   \parbox{24mm}{
    \begin{fmfchar*}(24,8)
    \fmfforce{(0.1w,0.1h)}{i1}
    \fmfforce{(0.9w,0.1h)}{o1}
    \fmfforce{(0.3w,0.8h)}{o2}
    \fmfforce{(0.5w,0.8h)}{o3}
    \fmfforce{(0.7w,0.8h)}{o4}
    \fmfforce{(0.3w,0.1h)}{v1}
    \fmfforce{(0.5w,0.1h)}{v2}
    \fmfforce{(0.7w,0.1h)}{v3}
    \fmf{fermion}{o1,v3,v2,v1,i1}
    \fmf{photon}{v1,o2}
    \fmf{photon}{v2,o3}
    \fmf{photon}{v3,o4}
    \fmfv{decor.shape=circle,decor.filled=1,decor.size=10}{i1}
    \fmfv{decor.shape=circle,decor.filled=1,decor.size=10}{o1}
    \fmfv{decor.shape=circle,decor.filled=1,decor.size=10}{v1}
    \fmfv{decor.shape=circle,decor.filled=1,decor.size=10}{v2}
    \fmfv{decor.shape=circle,decor.filled=1,decor.size=10}{v3}
    \end{fmfchar*}}
 \\ 
\vdots \end{array}\right)
\end{eqnarray*}

In other words, $\phi^n(\tunn)$  is a propagator with $n$ dangling
photon lines ($n$ starts at 0). More generally, all the components 
$\phi^n(t^\circ)$ or $\phi^n(t^\bullet)$ has $n$ dangling photon lines.
To see the action of the recursive equation for the photon,
observe that, in Eq.(\ref{phiblanc}) the fermion extremities of $t_1^\bullet$
are closed on an additional photon line by a bare vertex (on the left), and 
each dangling photon line
of $t_1^\bullet$ is linked in turn to the photon extremities
of  $t_2^\circ$ (on the right).
Diagrammatically: \parbox{18mm}{\begin{fmfgraph}(30,12)
    \fmfforce{(0.1w,0.5h)}{l}
    \fmfforce{(0.25w,0.5h)}{i1}
    \fmfforce{(0.45w,0.5h)}{i2}
    \fmfforce{(0.75w,0.5h)}{i3}
    \fmfforce{(0.9w,0.5h)}{r}
  \fmf{photon}{l,i1}
  \fmf{fermion,right}{i1,i2}
  \fmf{fermion,right}{i2,i1}
  \fmf{photon}{i2,i3}
  \fmf{photon}{i3,r}
  \fmfv{decor.shape=circle,decor.filled=0,%
   label=\mbox{\small$4$},label.angle=0,label.dist=0,decor.size=10}{l}
    \fmfv{decor.shape=circle,decor.filled=1,decor.size=10,label=\mbox{$y$}}{r}
    \fmfv{decor.shape=circle,decor.filled=1,decor.size=10,label=\mbox{$z$}}{i1}
    \fmfv{decor.shape=circle,decor.filled=0.5,decor.size=100}{i2}
    \fmfv{decor.shape=circle,decor.filled=0.1,decor.size=90}{i3}
\end{fmfgraph}}

The next term is 

\begin{eqnarray*}
\phi(\ttroisb) &=& \left( \begin{array}{c} 
   \parbox{18mm}{\begin{fmfgraph}(18,12)
   \fmfleft{l}
   \fmfright{r}
  \fmf{photon,tension=4}{l,i1}
  \fmf{fermion,right}{i1,i2}
  \fmf{fermion,right}{i2,i1}
  \fmf{photon,tension=4}{i2,r}
    \fmfv{decor.shape=circle,decor.filled=1,decor.size=10}{l}
    \fmfv{decor.shape=circle,decor.filled=1,decor.size=10}{r}
    \fmfv{decor.shape=circle,decor.filled=1,decor.size=10}{i1}
    \fmfv{decor.shape=circle,decor.filled=1,decor.size=10}{i2}
\end{fmfgraph}}
 \\ 
   \parbox{20mm}{
    \begin{fmfchar*}(20,15)
    \fmfforce{(0.1w,0.1h)}{i1}
    \fmfforce{(0.9w,0.1h)}{i2}
    \fmfforce{(0.5w,0.8h)}{i3}
    \fmfforce{(0.35w,0.1h)}{v1}
    \fmfforce{(0.65w,0.1h)}{v2}
    \fmfforce{(0.5w,0.55h)}{v3}
    \fmf{fermion}{v1,v2,v3,v1}
    \fmf{photon}{i1,v1}
    \fmf{photon}{i2,v2}
    \fmf{photon}{i3,v3}
    \fmfv{decor.shape=circle,decor.filled=1,decor.size=10}{i1}
    \fmfv{decor.shape=circle,decor.filled=1,decor.size=10}{i2}
    \fmfv{decor.shape=circle,decor.filled=1,decor.size=10}{v1}
    \fmfv{decor.shape=circle,decor.filled=1,decor.size=10}{v2}
    \fmfv{decor.shape=circle,decor.filled=1,decor.size=10}{v3}
    \end{fmfchar*}}+
   \parbox{20mm}{
    \begin{fmfchar*}(20,15)
    \fmfforce{(0.1w,0.8h)}{i1}
    \fmfforce{(0.9w,0.8h)}{i2}
    \fmfforce{(0.5w,0.1h)}{i3}
    \fmfforce{(0.35w,0.8h)}{v1}
    \fmfforce{(0.65w,0.8h)}{v2}
    \fmfforce{(0.5w,0.45h)}{v3}
    \fmf{fermion}{v1,v3,v2,v1}
    \fmf{photon}{i1,v1}
    \fmf{photon}{i2,v2}
    \fmf{photon}{i3,v3}
    \fmfv{decor.shape=circle,decor.filled=1,decor.size=10}{i1}
    \fmfv{decor.shape=circle,decor.filled=1,decor.size=10}{i2}
    \fmfv{decor.shape=circle,decor.filled=1,decor.size=10}{v1}
    \fmfv{decor.shape=circle,decor.filled=1,decor.size=10}{v2}
    \fmfv{decor.shape=circle,decor.filled=1,decor.size=10}{v3}
    \end{fmfchar*}}
 \\ 
\vdots \end{array}\right)
\end{eqnarray*}

For the fermion propagator,
in Eqs.(\ref{phinoir}) and (\ref{phinoir2}),
each dangling photon line
of $t_1^\bullet$ is linked in turn to the right photon extremity
of  $t_2^\circ$.
The left photon extremity of  $t_2^\circ$ is linked to the
left extremity of the electron propagator $t_1^\bullet$ by 
an additional bare vertex. 

If we neglect the first term in
Eq.(\ref{phinoir2}) we can write
diagrammatically: \parbox{18mm}{\begin{fmfgraph}(40,10)
    \fmfforce{(0.1w,0.1h)}{l}
    \fmfforce{(0.3w,0.1h)}{o1}
    \fmfforce{(0.6w,0.1h)}{o3}
    \fmfforce{(0.9w,0.1h)}{r}
  \fmf{fermion}{r,o3,o1,l}
  \fmf{photon,left}{o1,o3}
    \fmfv{decor.shape=circle,decor.filled=1,decor.size=10}{l}
    \fmfv{decor.shape=circle,decor.filled=1,decor.size=10}{o1}
    \fmfv{decor.shape=circle,decor.filled=0.5,decor.size=100}{o3}
    \fmfv{decor.shape=circle,decor.filled=1,decor.size=10}{r}
  \fmfforce{(0.45w,0.75h)}{o2}
  \fmfv{decor.shape=circle,decor.filled=0.1,decor.size=90}{o2}
\end{fmfgraph}}

The next term is
\begin{eqnarray*}
\phi(\ttroisn) &=& \left( \begin{array}{c} 
   \parbox{28mm}{\begin{fmfgraph}(28,8)
    \fmfforce{(0.1w,0.8h)}{l}
    \fmfforce{(0.37w,0.8h)}{i1}
    \fmfforce{(0.63w,0.8h)}{i2}
    \fmfforce{(0.9w,0.8h)}{r}
  \fmf{fermion}{r,i2,i1,l}
  \fmffreeze
  \fmf{photon,left}{i2,i1}
    \fmfv{decor.shape=circle,decor.filled=1,decor.size=10}{l}
    \fmfv{decor.shape=circle,decor.filled=1,decor.size=10}{r}
    \fmfv{decor.shape=circle,decor.filled=1,decor.size=10}{i1}
    \fmfv{decor.shape=circle,decor.filled=1,decor.size=10}{i2}
\end{fmfgraph}}
 \\ 
   \parbox{30mm}{
    \begin{fmfchar*}(30,12)
    \fmfforce{(0.1w,0.4h)}{i1}
    \fmfforce{(0.3w,0.4h)}{i2}
    \fmfforce{(0.5w,0.4h)}{i3}
    \fmfforce{(0.7w,0.4h)}{i4}
    \fmfforce{(0.9w,0.4h)}{i5}
    \fmfforce{(0.3w,0.9h)}{o}
    \fmf{fermion}{i5,i4,i3,i2,i1}
    \fmf{photon,left}{i4,i3}
    \fmf{photon}{i2,o}
    \fmfv{decor.shape=circle,decor.filled=1,decor.size=10}{i1}
    \fmfv{decor.shape=circle,decor.filled=1,decor.size=10}{i2}
    \fmfv{decor.shape=circle,decor.filled=1,decor.size=10}{i3}
    \fmfv{decor.shape=circle,decor.filled=1,decor.size=10}{i4}
    \fmfv{decor.shape=circle,decor.filled=1,decor.size=10}{i5}
    \end{fmfchar*}}+
   \parbox{30mm}{
    \begin{fmfchar*}(30,12)
    \fmfforce{(0.1w,0.4h)}{i1}
    \fmfforce{(0.3w,0.4h)}{i2}
    \fmfforce{(0.5w,0.4h)}{i3}
    \fmfforce{(0.7w,0.4h)}{i4}
    \fmfforce{(0.9w,0.4h)}{i5}
    \fmfforce{(0.5w,0.9h)}{o}
    \fmf{fermion}{i5,i4,i3,i2,i1}
    \fmf{photon,left}{i4,i2}
    \fmf{photon}{i3,o}
    \fmfv{decor.shape=circle,decor.filled=1,decor.size=10}{i1}
    \fmfv{decor.shape=circle,decor.filled=1,decor.size=10}{i2}
    \fmfv{decor.shape=circle,decor.filled=1,decor.size=10}{i3}
    \fmfv{decor.shape=circle,decor.filled=1,decor.size=10}{i4}
    \fmfv{decor.shape=circle,decor.filled=1,decor.size=10}{i5}
    \end{fmfchar*}}+\\
   \parbox{30mm}{
    \begin{fmfchar*}(30,12)
    \fmfforce{(0.1w,0.4h)}{i1}
    \fmfforce{(0.3w,0.4h)}{i2}
    \fmfforce{(0.5w,0.4h)}{i3}
    \fmfforce{(0.7w,0.4h)}{i4}
    \fmfforce{(0.9w,0.4h)}{i5}
    \fmfforce{(0.7w,0.9h)}{o}
    \fmf{fermion}{i5,i4,i3,i2,i1}
    \fmf{photon,left}{i3,i2}
    \fmf{photon}{i4,o}
    \fmfv{decor.shape=circle,decor.filled=1,decor.size=10}{i1}
    \fmfv{decor.shape=circle,decor.filled=1,decor.size=10}{i2}
    \fmfv{decor.shape=circle,decor.filled=1,decor.size=10}{i3}
    \fmfv{decor.shape=circle,decor.filled=1,decor.size=10}{i4}
    \fmfv{decor.shape=circle,decor.filled=1,decor.size=10}{i5}
    \end{fmfchar*}}
 \\ 
\vdots \end{array}\right)
\end{eqnarray*}

\subsection{Diagram enumeration}

As a warm-up exercice, we can calculate the number of diagrams 
in the component $\phi^n(t)$, that we denote $|\phi^n(t)|$.
From Eqs.(\ref{phinoir}) and (\ref{phiblanc}) we find the
equations for $|\phi^n(t^\circ)|$ and $|\phi^n(t^\bullet)|$
\begin{eqnarray*}
|\phi^n(t^\bullet)|&=&\sum_{k=0}^n\sum_{k'=0}^{n-k}(k'+1)|\phi^k(t_1^\circ)|
        |\phi^{k'+1}(t_2^\bullet)|,\\ 
|\phi^n(t^\circ)|&=&\sum_{k=0}^n (n-k+1) |\phi^k(t_1^\circ)| 
   |\phi^{n-k+1}(t_2^\bullet)|,
\end{eqnarray*}
with $\phi^n(\tunn)=1$, $\phi^n(\tunb)=\delta_{n,0}$.

The solution of this recursive equation is
\begin{eqnarray*}
|\phi^n(t^\bullet)|&=& \bar{\varphi}(t) {2|t|+n \choose n},\\
|\phi^n(t^\circ)|&=&  \bar{\varphi}(t) {2|t|+n-1 \choose n},
\end{eqnarray*}
where $\bar{\varphi}(t)$ does not depend on the colour of $t$ and was defined
in section \ref{enumsection}.

This can be used to calculate the total number of diagrams for the
electron or photon propagators (the number is the same) contributing at $e^{2n}$.
Let us define
\begin{eqnarray*}
s_n & = & \sum_{t\in Y_n} \bar{\varphi}(t)=
   \sum_{t\in Y_n} \bar{\varphi}(t_1)(2|t_2|+1)\bar{\varphi}(t_2).
\end{eqnarray*}
Using Eq.(\ref{dectree}) for $n>0$ we find
\begin{eqnarray*}
s_n& = & \sum_{k=0}^{n-1}\sum_{|t_1|=k}\bar{\varphi}(t_1)
\sum_{|t_2|=n-k-1}  (2|t_2|+1)\bar{\varphi}(t_2)\\
& = & \sum_{k=0}^{n-1} (2k+1) s_k s_{n-k-1}.
\end{eqnarray*}
The starting value is $s_0=1$.
For $n=0,1,2,3,4,5$ this gives us $1,1,4,27,248,2830$, in agreement with
Refs.\cite{Touchard,Stein}. The generating function $y(x)$ 
for the sequence $s_n$
satisfies the differential equation $2x^2 y y'+x y^2 -y +1=0$ with $y(0)=1$.

This enumation takes into account neither symmetry nor Furry's theorem, 
which says that $|\phi^n(t^\circ)|=0$
if $n$ is odd.
The main point of this enumeration is to show that each tree 
represents the sum of a large 
number of diagrams when $|t|$ is large. 
This may prove useful for practical calculations.

\subsection{Fourier transform}
In applications, it is often convenient to work in the 
$k$-space.
The Fourier transform of $\phi(t;x,y;\{\lambda,z\}_{1,n})$
is defined by
\begin{eqnarray*}
\psi(t;q,q';\{\lambda,p\}_{1,n})&=& \int 
  d^4x d^4y d^4z_1 \dots d^4z_n\\&&\hspace*{-2cm}
  e^{i(q\cdot x-q'\cdot y+p_1\cdot z_1+\cdots+p_n\cdot z_n)}
  \phi(t;x,y;\{\lambda,z\}_{1,n}).
\end{eqnarray*}
This corresponds to outgoing momenta $p_i$ along the dangling photon lines.
If this is introduced into Eqs.(\ref{phinoir}) and (\ref{phiblanc}),
we find 
\begin{eqnarray*}
\psi(t;q,q';\{\lambda,p\}_{1,n})&=& (2\pi)^4
  \delta(q+p_1+\cdots+p_n-q') \\&&\hspace*{10mm}
  \tilde\phi(t;q;\{\lambda,p\}_{1,n}).
\end{eqnarray*}

The full fermion and photon propagators in Fourier space are
\begin{eqnarray*}
S(q)&=&\sum_t e^{2|t|}\tilde\phi^0(t^\bullet;q),\\
D_{\lambda\mu}(q)&=&\sum_t e^{2|t|}\tilde\phi^0_{\lambda\mu}(t^\circ;q).
\end{eqnarray*}

Here $\tilde\phi(t)$ satisfies the recursive relation
\begin{eqnarray}
\tilde{\phi}^n(t^\bullet;q;\{\lambda,p\}_{1,n})&=& 
 S^0(q)\gamma^{\lambda_1}\nonumber\\&&\hspace*{-20mm}
 \tilde{\phi}^{n-1}(t^\bullet;q+p_1;\{\lambda,p\}_{2,n})+
  \nonumber\\&&\hspace*{-23mm}
 i\sum_{k=0}^{n}\int \frac{d^4 p}{(2\pi)^4} S^0(q)\gamma^\lambda
 \tilde{\phi}^k_{\lambda\lambda'} (t_1^\circ;p;\{\lambda,p\}_{1,k})
  \nonumber\\&&\hspace*{-20mm}
 \tilde{\phi}^{n-k+1}_\Sigma(t_2^\bullet;q-p;\lambda',p+P_k,
  \{\lambda,p\}_{k+1,n}),\label{phinoirk}\\
\tilde{\phi}^n_{\mu\nu}(t^\circ;q;\{\lambda, p\}_{1,n})&=& 
 -i\sum_{k=0}^{n}\int \frac{d^4p}{(2\pi)^4} D^0_{\mu\lambda}(q)
  \nonumber\\&&\hspace*{-29mm}
 \tilde{\phi}^{k}_{\lambda'\nu}(t_1^\circ;q+P_{k};\{\lambda, p\}_{1,k})
  \nonumber\\&&\hspace*{-29mm}
  \mathrm{tr}\big[\gamma^\lambda
  \tilde{\phi}^{n-k+1}_\Sigma(t_2^\bullet;p;\lambda',-q-P_{k},
   \{\lambda, p\}_{k+1,n})
  \big], \label{phiblanck}
\end{eqnarray}
where we have noted $P_k=p_1+\cdots+p_k$ and 
with the initial data
\begin{eqnarray*}
\tilde{\phi}^0_{\mu\nu}(\tunb;q) &=& D^0_{\mu\nu}(q)=
-\frac{g_{\mu\nu}-(1-1/\xi)q_\mu q_\nu/q^2}{q^2+i\epsilon},\\
\tilde{\phi}^n_{\mu\nu}(\tunb;q) &=& 0\quad\mathrm{for}\quad n\ge 1,\\
\tilde{\phi}^0(\tunn;q) &=& S^0(q)={(\gamma^\mu q_\mu-m+i\epsilon)}^{-1},\\
\tilde{\phi}^1(\tunn;q;\lambda_1,p_1) &=& S^0(q)\gamma^{\lambda_1}S^0(q+p_1),\\
\tilde{\phi}^n(\tunn;q;\{\lambda,p\}_{1,n}) &=& S^0(q)\gamma^{\lambda_1}
  S^0(q+p_1)\gamma^{\lambda_2}\cdots\gamma^{\lambda_n}\\&&
  \hspace*{15mm}
  S^0(q+p_1+\cdots+p_n).
\end{eqnarray*}

As for the real space case, the first term of Eq.(\ref{phinoirk})
is absent for $n=0$ and 
\begin{eqnarray*}
\tilde{\phi}_\Sigma^n(t;q;\{\lambda,p\}_{1,n}) &=& 
\tilde{\phi}^n(t;q;\{\lambda,p\}_{1,n}) +\\&&\hspace*{-10mm}
\tilde{\phi}^n(t;q;\lambda_2,p_2,\lambda_1,p_1,\{\lambda,p\}_{3,n}) +
\cdots+\\&&\hspace*{-10mm}
\tilde{\phi}^n(t;q;\{\lambda,p\}_{2,n},\lambda_1,p_1).
\end{eqnarray*}
Again, Furry's theorem enables us to restrict the sum to the
even components of $\tilde{\phi}_{\mu\nu}(t^\circ)$.

\section{Renormalization}

In Ref.\cite{Brouder}, it has been shown that the renormalization
through Hopf algebra created by Kreimer could be done also within 
the Butcher approach using a difference of Runge-Kutta methods
(bare quantities minus counterterms). 
The same method can be tried here.

If $X=A+F(X,\frac{\delta X}{\delta v(z)})$ is the equation to
renormalize, and if $R$ represents a renormalization prescription 
(see Ref.\cite{Kreimer}),
then the renormalized field $X_R$ is given by
\begin{eqnarray*}
X_R&=& \sum_{t} \phi_R^0 (t),
\end{eqnarray*}
where $\phi_R (t)$ is defined recursively from the integral operator 
$F-R\circ F$ by $\phi^n_R(\tunn) = \phi^n(\tunn)$ and
\begin{eqnarray*}
\phi_R^n(t;\{z\}_{1,n}) &=& \sum_{k=0}^n\\&&\hspace*{-20mm}
  F\big(\phi_R^k(t_1;\{z\}_{1,k}),
  \phi^{n-k+1}_{R\Sigma}(t_2;z,\{z\}_{k+1,n}))-\\&&\hspace*{-20mm}
  R\Big[F\big(\phi_R^k(t_1;\{z\}_{1,k}),
  \phi^{n-k+1}_{R\Sigma}(t_2;z,\{z\}_{k+1,n})\big)\Big].
\end{eqnarray*}

Within this approach, renormalization 
looks rather simple. 
At each stage of a calculation,
use the renormalized $\phi$ instead of the bare ones,
and subtract the singularity. 
This rule seems promissing for the case of QED, where $\phi^n(\tunn)$
is regular and needs not be renormalized, and where the
recursive operations described by Eqs.(\ref{phinoirk}) and (\ref{phiblanck})
involve only a single loop (a fermion loop
for the photon propagator and a photon loop for the fermion
propagator). Since $\phi^k_R(t_1)$ and $\phi^{n-k+1}_R(t_2)$ are 
regular, there is no subdivergence, and thus no
overlapping divergence problem.

Following Kreimer \cite{Kreimer}, the renormalization group 
equations are a consequence 
of the fact that $F-R(F)=R'(F)-R(F)+ F-R'(F)$ is independent of
the parameters of $R'$.

Clearly, this presentation of renormalization is 
far too sketchy and 
further investigation is required to include renormalization 
prescriptions and calculation of counterterms, and 
to check that the
results are the same as in the standard theory.

\section{Applications}
In this section, the previous results are applied to Furry's
theorem and the Ward-Takahashi identities. Finally,
a sum over trees is defined for the two-particle 
Green function.

\subsection{Furry's theorem \label{Furrysection}}
Within our approach, Furry's theorem implies 
$\phi^{n}(t^\circ)=0$ for odd n. To show this, remark that, in 
Eq.(\ref{phiblanck}), 
the integral
\begin{eqnarray*}
 \int \frac{d^4p}{(2\pi)^4} 
  \mathrm{tr}\big[\gamma^\lambda
  \tilde{\phi}^{n-k+1}_\Sigma(t_2^\bullet;p;\lambda',-q-P_{k},
   \{\lambda, p\}_{k+1,n} \big]
\end{eqnarray*}
is a fermion loop with $n-k+2$ external photon lines. According to 
Furry's theorem,
this loop is zero when $n-k$ is odd. 
From its explicit definition, $\phi^{n}(\tunb)=0$ is zero if $n$ is odd 
(in fact,
it is zero if $n\ge 1$). We reason recursively on the number of 
vertices of $t^\circ$.
If $\phi^{n}(t^\circ)=0$ for odd $n$ and $t^\circ$ with up to $2N-1$ vertices,
take $t^\circ$ with  $2N+1$ vertices. In Eq.(\ref{phiblanck}), the integral 
over $p$ is zero
if $n-k$ is odd (Furry's theorem) and $\phi^{k}(t_1^\circ)$ is zero if 
$k$ is odd
(because $t_1^\circ$ has less vertices than $t$). Thus, for the left-hand 
side of 
Eq.(\ref{phiblanck}) to be eventually non-zero, $k$ and $n-k$ must be even, 
so $n$
must be even. Therefore, $\phi^{n}(t^\circ)=0$ for odd $n$.

\subsection{Ward-Takahashi identities}
Within the present approach, the Ward-Takahashi identities take the form
\begin{eqnarray}
\sum_\mu k_\mu\tilde{\phi}^{n+1}_\Sigma(t^\bullet;q;
   \mu,k,\{\lambda,p\}_{1,n}) &=&
\tilde{\phi}^{n}(t^\bullet;q;\{\lambda,p\}_{1,n}) -\nonumber\\&&
  \hspace*{-10mm}
\tilde{\phi}^{n}(t^\bullet;q+k;\{\lambda,p\}_{1,n}). \label{WT}
\end{eqnarray}

Again, these identities will be proved recursively over the number of 
vertices of 
$t^\bullet$. An elementary calculation shows that Eq.(\ref{WT}) is true
for $t^\bullet=\tunn$ (Ref.\cite{Peskin}, p.238). 
Using Eq.(\ref{recure}), we can write Eq.(\ref{WT}) as
\begin{eqnarray}
\sum_\mu k_\mu\frac{\delta\tilde{\phi}^{n}(t^\bullet;q;\{\lambda,p\}_{1,n})}
   {\delta A_\mu(k)} &=&
\tilde{\phi}^{n}(t^\bullet;q;\{\lambda,p\}_{1,n}) -
\nonumber\\&&\hspace*{-10mm}
\tilde{\phi}^{n}(t^\bullet;q+k;\{\lambda,p\}_{1,n}). \label{WT2}
\end{eqnarray}
The last equation can be proved by applying the differential operator
$\sum_\mu k_\mu\frac{\delta}{\delta A_\mu(k)}$ to both sides of 
Eq.(\ref{phinoirk}).
Consider first the second term of Eq.(\ref{phinoirk}). The differential
operator acting on $\tilde{\phi}^k_{\lambda\lambda'}(t^\circ_1)$ adds
a photon line to $\tilde{\phi}^k_{\lambda\lambda'}(t^\circ_1)$, which gives
zero by Furry's theorem. The differential operator acting on 
$\tilde{\phi}^{n-k+1}_{\Sigma}(t^\bullet_2)$ will transform according to
Eq.(\ref{WT2}), which is valid because $t^\bullet_2$ has less vertices than
$t^\bullet$. Carrying out the sum over $k$, we see that Eq.(\ref{WT2}) is
true for the second term of the right-hand side of Eq.(\ref{phinoirk}).

Consider now the first term of Eq.(\ref{phinoirk}). For $n=0$ it does
not exist, so Eq.(\ref{WT2}) is proved for $n=0$ because it is true
for the second term. The recursion runs now over $n$, the 
functional derivative operates 
either on $S^0(q)$ or on
$\tilde{\phi}^{n-1}(t^\bullet;q+p_1;\{\lambda,p\}_{2,n})$. Since
Eq.(\ref{WT2}) is true for both, and for the second term, it is
true for $\tilde{\phi}^{n}(t^\bullet;q;\{\lambda,p\}_{1,n})$.

\subsection{Homology and Ward-Takahashi identities}
The previous Ward-Takashi indentities induce a boundary
operator on $\phi(t)$. In that case, we need to start 
from the fully symmetrized form of $\tilde{\phi}$:
\begin{eqnarray*}
\tilde{\phi}^n_{\Sigma^n}(t;q;\{\lambda,z\}_{1,n})=
\sum_{\sigma\in \mathcal{S}_n}
  \tilde{\phi}^n (t;q;\{\lambda,z\}_{\sigma(1),\sigma(n)}).
\end{eqnarray*}

We define $\partial_i$ by
\begin{eqnarray*}
\partial_i \tilde{\phi}^n_{\Sigma^n}(t;q;\{\lambda,z\}_{1,i-1},
 \mu,k,\{\lambda,z\}_{i+1,n})&=&\\
\sum_{\mu=0}^3 k_\mu \tilde{\phi}^n_{\Sigma^n}(t;q;\{\lambda,z\}_{1,i-1},
 \mu,k,\{\lambda,z\}_{i+1,n}) &=&\\&&\hspace*{-50mm}
 \tilde{\phi}^{n-1}_{\Sigma^{n-1}}(t;q;\{\lambda,z\}_{1,i-1},
 \{\lambda,z\}_{i+1,n})-\\&&\hspace*{-50mm}
  \tilde{\phi}^{n-1}_{\Sigma^{n-1}}(t;q+k;\{\lambda,z\}_{1,i-1},
 \{\lambda,z\}_{i+1,n}),
\end{eqnarray*}
where the last line is the Ward-Takahashi identity.
If we write the usual expression for the boundary operator
\begin{eqnarray*}
\partial \tilde{\phi}^n_{\Sigma^n}(t;q;\{\lambda,z\}_{1,n})=\sum_{i=1}^n
(-1)^i \partial_i \tilde{\phi}^n_{\Sigma^n}(t;q;\{\lambda,z\}_{1,n}),
\end{eqnarray*}
we find by explicit calculation that $\partial^2=0$.
This was proved for the fermion component, but 
$\partial^2=0$ for the photon component too, because
of Furry's theorem. Notice that,
for $n=0$ we have put $\partial\phi^0(t)=0$.

\subsection{Two-particle Green function}
In Ref.\cite{Schwinger} Schwinger proceeds by giving an
equation for the full two-particle fermion Green function:
\begin{eqnarray*}
&& \Big[i\gamma^\mu\partial_{x_1^\mu}-m-e\gamma^\mu A_\mu(x_1;J)
 +ie\gamma^\mu \frac{\delta}{\delta J_\mu(x_1)} \Big] \nonumber\\&&
  \hspace*{8mm}
   S(x_1,x_2;y_1,y_2;J)
   =\delta(x_1-y_1)S(x_2,y_2;J)-
 \nonumber\\&&
  \hspace*{18mm}
\delta(x_1-y_2)S(x_2,y_1;J). 
\end{eqnarray*}
From the identity \cite{Bogoliubov} 
\begin{eqnarray*}
\frac{\delta}{\delta J^\mu(x)}=-\int dy 
G_{\lambda\mu}(y,x)
\frac{\delta}{\delta A_\lambda(y)},
\end{eqnarray*}
we go from the $J$ to the $A$ variable (in the limit $A$=0)
\begin{eqnarray*}
&&\Big[i\gamma^\mu\partial_{x_1^\mu}-m-e\gamma^\mu A_\mu(x_1)]
   S(x_1,x_2;y_1,y_2;A)=\\&&\hspace*{2mm} 
   \delta(x_1-y_1)S(x_2,y_2;A)-\delta(x_1-y_2)S(x_2,y_1;A)+\\&&
 \hspace*{3mm}
  ie\int d^4z \gamma^\mu \frac{\delta S(x_1,x_2;y_1,y_2;A)}
 {\delta A_\lambda(z)} G_{\lambda\mu}(z,x_1).
\end{eqnarray*}
It remains to multiply by $S^0$ and to take $A=0$.
This yields
\begin{eqnarray}
S(x_1,x_2;y_1,y_2)&=&S^0(x_1,y_1)S(x_2,y_2)-S^0(x_1,y_2)S(x_2,y_1)
  \nonumber\\ &&\hspace*{-26mm}
 +ie\int d^4z d^4z' S^0(x_1,z) \gamma^\mu 
  \frac{\delta S(z,x_2;y_1,y_2)}
 {\delta A_\lambda(z')} G_{\lambda\mu}(z',z). \label{twobody}
\end{eqnarray}

In Eq.(\ref{twobody}), $S(x,y)$ plays the role of initial data,
whereas it is the solution of Eq.(\ref{eqbase}). Therefore, we are
making a composition of solutions, which was considered in 
section \ref{groupsection}.
The situation is not exactly the same as that of 
section \ref{groupsection}, but the proof is similar
(only more cumbersome) and we obtain an expression for
the two-particule Green function as a sum over planar
binary trees
\begin{eqnarray*}
S(x_1,x_2;y_1,y_2)&=&\sum_t e^{2|t|}
\chi^0(t;x_1,x_2;y_1,y_2).
\end{eqnarray*}
According to the rule of composition, $\chi$ is the sum of
two terms:
\begin{eqnarray*}
\chi^n(t;x_1,x_2;y_1,y_2;\{\lambda,z\}_{1,n}) &=&\\&&\hspace*{-30mm}
S^0(x_1,y_1)\phi^0(t^\bullet;x_2,y_2;\{\lambda,z\}_{1,n})-\\&&\hspace*{-30mm}
S^0(x_1,y_2)\phi^0(t^\bullet;x_2,y_1;\{\lambda,z\}_{1,n})+\\&&\hspace*{-30mm}
\psi^0(t;x_1,x_2;y_1,y_2;\{\lambda,z\}_{1,n}),
\end{eqnarray*}
and $\psi$ itself is given by 
\begin{eqnarray*}
\psi^n(t;x_1,x_2;y_1,y_2;\{\lambda,z\}_{1,n})&=&
S^0(x_1,z_1)\gamma^{\lambda_1}\\&&\hspace{-35mm}
  \chi^{n-1}(t;z_1,x_2;y_1,y_2;\{\lambda,z\}_{2,n})+\\&&\hspace{-40mm}
  i\sum_{k=0}^n \int d^4z d^4z' S^0(x_1,z)\gamma^\mu
    \phi^{k}_{\lambda\mu}(t_1^\circ;z',z;\{\lambda,z\}_{1,k})
   \\&&\hspace{-40mm}
 \chi^{n-k+1}_\Sigma(t_2;z,x_2;y_1,y_2;\lambda,z',\{\lambda,z\}_{k+1,n}).
 \label{twopartgreen}
\end{eqnarray*}

Equation (\ref{twopartgreen}) is not very simple, but it provides
a way to calculate recursively all orders of the perturbation
expansion for the two-particle Green function of QED. In that sense,
it is not so complicated.

\section{Vacuum polarization, self-energy}

To calculate vacuum polarization and self-energy, we introduce
a further operation on planar binary trees.
Let 
\begin{eqnarray}
P(t)=\sum_{i=1}^{n(t)} u_i \otimes v_i, \label{Pdet}
\end{eqnarray}
where $n(t)$ is an integer recursively defined by
\begin{eqnarray*}
n(\tunn) &= & 0, \\
n(t) &= & 0\quad\mathrm{if}\quad t=t_1\vee\tunn, \\
n(t) &= & 1+n(t_2) \quad\mathrm{if}\quad t=t_1\vee t_2,\quad t_2\not=\tunn,
\end{eqnarray*}
and the planar binary trees $u_i$ and $v_i$ are determined by
\begin{eqnarray}
P(\tunn) &= & 0, \nonumber\\
P(t) &= & 0\quad\mathrm{if}\quad t=t_1\vee\tunn, \nonumber\\
P(t) &= & (t_1\vee \tunn)\otimes t_2+
  \sum_{i=1}^{n(t_2)} (t_1\vee u_i)\otimes v_i\nonumber\\&&\hspace*{8mm}
 \quad\mathrm{if}\quad t=t_1\vee t_2,\quad t_2\not=\tunn.
 \label{defP}
\end{eqnarray}
The trees $u_i$ and $v_i$ in Eq.(\ref{defP}) are generated by
Eq.(\ref{Pdet}) for $t=t_2$.

As a more graphical definition, for a tree $t$, we consider
the path starting from the root and climbing up the tree by
taking, at each vertex, the right branch. This path terminates
at the extreme right leaf of the tree and goes through
$n(t)+2$ vertices (including the root and the leaf). For each
vertex $s_i$ along that path, excluding the root and the leaf,
we cut $t$ into two trees $u_i$ and $v_i$, where
$v_i$ is the subtree of $t$ that has $s_i$ as a root, and
$u_i$ the subtree of $t$ that has $s_i$ as a leaf.
For example:
\begin{eqnarray*}
P\left(\tcinqnn\quad\right)= \tquatredeuxnn\otimes\ttroisnn .
\end{eqnarray*}

This operation will help calculating the vacuum polarization
and the self-energy.
If we define $Y$ by
\begin{eqnarray*}
X=\sum_t \phi^0(t)= \frac{1}{1/\phi^0(\tunn) -Y},
\end{eqnarray*}
it is proved in the appendix that $Y=\sum_t \psi^0(t)$, where
\begin{eqnarray}
\psi^0(\tunn)&=& 0,\nonumber\\
\psi^0(t)&=& \frac{1}{\phi^0(\tunn)}\Big( \phi^0(t) \frac{1}{\phi^0(\tunn)}-
   \sum_{i=1}^{n(t)} \phi^0(u_i)\psi^0(v_i)\Big).
  \label{inverse}
\end{eqnarray}
In this equation, $u_i$, $v_i$ and $n(t)$ are
determined from $t$ by Eq.(\ref{Pdet}).

\subsection{Vacuum polarization}
From the general formula, we deduce that the vacuum polarization is
given, in terms of the map $\phi$ for the full photon propagator,
by
\begin{eqnarray*}
\Pi_{\lambda\mu}(q) &=& \sum_t e^{2|t|}\psi^0_{\lambda\mu}(t^\circ;q),
\end{eqnarray*}
with $\psi^0(\tunb;q)  =  0$ and
\begin{eqnarray*}
\psi^0_{\lambda\mu}(t^\circ;q) & = & (\gamma^\alpha q_\alpha -m) 
       \phi^0_{\lambda\mu}(t^\circ;q) (\gamma^\beta q_\beta -m) -
\nonumber\\&&
   (\gamma^\alpha q_\alpha -m) \sum_{i=1}^{n(t)} 
   \phi^0_{\lambda\nu}(u_i^\circ;q)
    g^{\nu\nu'} \psi^0_{\nu'\mu}(v_i^\circ;q).
\end{eqnarray*}

Incidentally, it can be noted that 
the full vacuum polarization 
$\Pi^{\lambda\mu}(q)$ is
related to the fermion loop 
by (Ref.\cite{Itzykson} p.477):
\begin{eqnarray*}
\Pi^{\lambda\mu}(q)=-i
 \sum_t e^{2|t|+2} \int \frac{d^4p}{(2\pi)^4} 
  \mathrm{tr}\big[\gamma^\lambda
  \tilde{\phi}^{1}(t^\bullet;p;\mu,-q)\big].
\end{eqnarray*}

\subsection{Self-energy}
Similarly, the self-energy is
given,
in terms of the map $\phi$ for the full fermion propagator,
by
\begin{eqnarray*}
\Sigma(q) &=& \sum_t e^{2|t|}\psi^0(t^\bullet;q),
\end{eqnarray*}
with $\psi^0(\tunn;q)  =  0$ and
\begin{eqnarray*}
\psi^0(t^\bullet;q) & = & (\gamma^\alpha q_\alpha -m) 
  \phi^0(t^\bullet;q) (\gamma^\beta q_\beta -m) -
\nonumber\\&&
   (\gamma^\alpha q_\alpha -m) 
  \sum_{i=1}^{n(t)} \phi^0(u_i^\bullet;q)\psi^0(v_i^\bullet;q).
\end{eqnarray*}

From the formula for the self-energy, we can deduce the
complete one-particle irreducible three-point function
$\Gamma^\nu(p,p')$ \cite{Itzykson}, p.335. 
We reintroduce the external potential to write
\begin{eqnarray*}
\Sigma(x,y;A)&=& \sum_t (i\gamma^\lambda \partial_{x^\lambda} -m - 
  e\gamma^\lambda A_\lambda(x))\\&&
  \phi(t^\bullet;x,y;A)
  (-i\gamma^\mu \overleftarrow{\partial}_{y^\mu} -m - e\gamma^\mu A_\mu(y)) -
  \\&&
  \sum_t (i\gamma^\lambda \partial_{x^\lambda} -m - 
  e\gamma^\lambda A_\lambda(x))\\&&
  \sum_{i=1}^{n(t)} \int dz'
  \phi(u_i^\bullet;x,z';A)\psi(v_i^\bullet;z',y;A),
\end{eqnarray*}
where $\overleftarrow{\partial}_{y^\mu}$ acts on the left.
In the real space, $\Gamma^{\nu}(x,y;z)$ is given 
\begin{eqnarray*}
\Gamma^{\nu}(x,y;z)=\frac{\delta \Sigma(x,y;A)}{\delta A_\nu(z)}
\quad\mathrm{for}\quad A=0.
\end{eqnarray*}
Therefore,
\begin{eqnarray*}
\Gamma^{\nu}(x,y;z) &=& \sum_t e^{|t|+1} \psi^1(t^\bullet;x,y;\nu,z),
\end{eqnarray*}
with 
\begin{eqnarray*}
\psi^1(t^\bullet;x,y;\nu,z) &=& -\gamma^\nu \delta(z-x)\\&&\hspace*{-10mm}
  \phi^0(t^\bullet;x,y)
  (-i\gamma^\mu \overleftarrow{\partial}_{y^\mu} -m)
  - \\&&\hspace*{-22mm}
  (i\gamma^\lambda \partial_{x^\lambda} -m) 
  \phi^0(t^\bullet;x,y)\gamma^\nu\delta(z-y)
  + \\&&\hspace*{-22mm}
  (i\gamma^\lambda \partial_{x^\lambda} -m) 
  \phi^1(t^\bullet;x,y;\nu,z)
  (-i\gamma^\mu \overleftarrow{\partial}_{y^\mu} -m )
  + \\&&\hspace*{-22mm}
  \gamma^\nu \delta(z-x) 
  \sum_{i=1}^{n(t)} \int dz'
  \phi^0(u_i^\bullet;x,z')\psi^0(v_i^\bullet;z',y)- \\&&\hspace*{-22mm}
  (i\gamma^\lambda \partial_{x^\lambda} -m )
  \sum_{i=1}^{n(t)} \int dz'
  \phi^0(u_i^\bullet;x,z')\psi^1(v_i^\bullet;z',y;\nu,z)-\\&&\hspace*{-22mm}
  (i\gamma^\lambda \partial_{x^\lambda} -m )
  \sum_{i=1}^{n(t)} \int dz'
  \phi^1(u_i^\bullet;x,z';\nu,z)\psi^0(v_i^\bullet;z',y).
\end{eqnarray*}
In Fourier space, this gives us
\begin{eqnarray*}
\Gamma^{\nu}(q,q+p) &=& \sum_t e^{|t|+1} \psi^1(t^\bullet;q;\nu,p),
\end{eqnarray*}
with 
\begin{eqnarray*}
\psi^1(t^\bullet;q;\nu,p) &=& -\gamma^\nu 
 \phi^0(t^\bullet;q)(i\gamma^\mu (q_\mu+p_\mu) -m)
 - \\&&\hspace*{-10mm}
 (i\gamma^\lambda q_\lambda -m) 
 \phi^0(t^\bullet;q) \gamma^\nu
 + \\&&\hspace*{-10mm}
 (i\gamma^\lambda q_\lambda -m) 
 \phi^1(t^\bullet;q;\nu,p)
 (i\gamma^\mu (q_\mu+p_\mu) -m )
 + \\&&\hspace*{-10mm}
 \gamma^\nu 
 \sum_{i=1}^{n(t)} 
 \phi^0(u_i^\bullet;q+p)\psi^0(v_i^\bullet;q+p)- \\&&\hspace*{-10mm}
 (i\gamma^\lambda q_\lambda -m )
 \sum_{i=1}^{n(t)} 
 \phi^0(u_i^\bullet;q)\psi^1(v_i^\bullet;q;\nu,p)  - \\&&\hspace*{-10mm}
 (i\gamma^\lambda q_\lambda -m )
 \sum_{i=1}^{n(t)} 
 \phi^1(u_i^\bullet;q;\nu,p)\psi^0(v_i^\bullet;q+p).
\end{eqnarray*}

\section{Interaction with an external field \label{extsection}}
In this section we come back to the original Schwinger equation,
because the presence of an external source is a convenient way
to represent the nuclei in the QED of matter.

Starting from Eq.(\ref{Schwinger}), we multiply by the corresponding
bare Green functions and we introduce $A^{\nu}(x;J)$ into the
second equation to obtain
\begin{eqnarray}
S(x,y;J) &= & S^0(x,y) -e\int d^4z d^4z' S^0(x,z) \gamma^\mu 
   D^0_{\mu\nu}(z,z') \nonumber\\&&\hspace*{-5mm} J^\nu(z') S(z,y;J)-
  ie^2\int d^4z d^4z' S^0(x,z) \gamma^\mu \nonumber\\&&\hspace*{-5mm}
   D^0_{\mu\nu}(z,z') \mathrm{tr}\big[ \gamma^\nu S(z',z';J) \big] 
 S(z,y;J)-
  \nonumber\\&&\hspace*{-5mm}
  ie\int d^4z S^0(x,z)  \gamma^\mu  
  \frac{\delta S(z,y;J)}{\delta J^\mu(z)}.
  \label{tadpole}
\end{eqnarray}

Here, the Schwinger equation is a sum of three terms. 
The first term is simply
the classical interaction with the external source $J^\nu(z_2)$, it
can be solved by defining a bare propagator in the
presence of this source
\begin{eqnarray*}
S^0(x,y;J)^{-1}=i\gamma^\mu\partial_\mu-m+e\gamma^\mu
\int d^4z D^0_{\mu\nu}(x,z) J^\nu(z).
\end{eqnarray*}
Equation (\ref{tadpole}) becomes now
\begin{eqnarray}
S(x,y;J) &= & S^0(x,y;J)-
  ie^2\int d^4z d^4z' S^0(x,z;J) \gamma^\mu \nonumber\\&&
   D^0_{\mu\nu}(z,z') 
   \mathrm{tr}\big[ \gamma^\nu S(z',z';J) \big] S(z,y;J)-\nonumber\\&&
  ie\int d^4z S^0(x,z;J)  
  \gamma^\mu  \frac{\delta S(z,y;J)}{\delta J^\mu(z)}.
  \label{tadpole2}
\end{eqnarray}

This equation is solved by the usual methods, and the recursive
definition of $\phi(t)$ is
\begin{eqnarray*}
\phi^n(\tunn;x,y;\{\lambda,z\}_{1,n})&=& (-1)^n \int d^4s_1\dots d^4s_n 
S^0(x,s_1;J) \\&&
\hspace*{-8mm}
\gamma^{\mu_1} D^0_{\mu_1\lambda_1}(s_1,z_1)S^0(s_1,s_2;J)\dots\\&&
\hspace*{-8mm}
\gamma^{\mu_n} D^0_{\mu_n\lambda_n}(s_n,z_n)S^0(s_n,y;J)\\
\phi^n(t;x,y;\{\lambda,z\}_{1,n})&=& -i 
 \sum_{k=0}^{n}\sum_{k'=0}^{n-k}\int d^4z d^4z' \\&&\hspace*{-3cm}
 \phi^{n-k-k'}(\tunn;x,z;\{\lambda, z\}_{1,n-k-k'})
\gamma^{\mu} D^0_{\mu\nu}(z,z')
  \\&&\hspace*{-25mm}\mathrm{tr}\big[\gamma^\nu
 \phi^k(t_1;z',z';\{\lambda,z\}_{n-k-k'+1,n-k'})\big]\\&&\hspace*{-25mm}
 \phi^{k'}(t_2;z,y;\{\lambda,z\}_{n-k'+1,n})-
 \\&&\hspace*{-3cm}i\sum_{k=0}^{n}\int d^4z
  \phi^{k}(\tunn;x,z;\{\lambda, z\}_{1,k})\nonumber\\&&\hspace*{-25mm}
  \gamma_\mu\phi^{n-k+1}_\Sigma(t_2;z,y;\mu,z,\{\lambda,z\}_{k+1,n}),
\end{eqnarray*}
where the last term is non zero only if $t$ has the special shape
$t=\tunn\vee t_2$.

According to Schwinger \cite{Schwinger}, the full photon Green function
in the presence of an external current $J$ is given by the functional
derivative of $A(x;J)$ with respect to $J(y)$. Therefore
\begin{eqnarray*}
D_{\lambda\mu}(x,y;J)&=&-\frac{\delta A_\lambda(x;J)}{\delta J^\mu(y)}\\
&=&D^0_{\lambda\mu}(x,y)+\\&&\hspace*{0mm}
         ie\int dz D^0_{\lambda\nu}(x,z)
           \mathrm{tr}\big[\gamma^\nu 
          \frac{\delta S(z,z;J)}{\delta J^\mu(y)}\big]\\
&=&D^0_{\lambda\mu}(x,y)+\\&&\hspace*{0mm}ie^2\sum_t\int dz 
          D^0_{\lambda\nu}(x,z)
           \mathrm{tr}\big[\gamma^\nu \phi^1(t;z,z;\mu,y)\big].
\end{eqnarray*}

It is also possible to start directly from Eq.(\ref{tadpole}) and to
write a tree solution of this equation using bare fermion Green functions.
Here, the strong field case was treated because it is probably more
interesting for applications to solid-state physics.

\section{Conclusion}
A method was presented to write the solution of some Schwinger equations
as a series over planar binary trees. 
In quantum field theory, it is common to expand over the number of loops
or to use integral equations relating, for instance, the full propagator
to the full vertex. The first method gives explicit results but becomes
very complex, and hundreds of diagrams must be built and calculated after
the first few terms of the perturbation expansion. The
second method is formally powerful but not very explicit because 
an $n$-body Green function is expressed in terms of an unknown 
($n$+$1$)-body Green function.
The present approach is a way to mix these two methods to obtain an
explicit recursive formula for the propagators and its functional 
derivatives.

The main point of the method is that explicit recursive expressions
can be given for the solution of Schwinger equations. 
Because of the recursive structure, the results
obtained at each step can be reused for the next step. 

The present paper is only
a first exploration of the method of series indexed
by planar binary trees, and much work remains to be done to explore its 
algebraic properties and its applications.

Two kinds of applications were presented here. On the one hand, a
series indexed by planar binary trees was given for various
physical quantities (full fermion and photon propagators, 
full two-body Green function), and a method was given to deduce
from this a series for vacuum polarization, fermion self-energy and
irreducible vertex function. Although the formulas may be a bit
cumbersome, they are derived and proved easily.
On the other hand, the recursive nature of the terms of the series
is well suited to prove properties
to all orders of perturbation theory. This was done in the
case of the Ward-Takahashi identities.

The present work can be expanded in various directions. Other field theories
can be investigated, as well as many-particle Green functions. The present
treatment was restricted to classical electromagnetic sources, it could be
worthwile to study the case of anticommuting fermion sources.
Furthermore, planar binary trees could be used to solve the
Hedin equation \cite{Hedin} and investigate the 
GW-approximation \cite{GW} of solid-state physics.

As a last point, it can be noticed that previous articles
have presented general planar trees as the structure
adapted to quantum field theory \cite{ConnesK}.
In fact, planar trees and planar binary trees are
equivalent for that purpose.
Since the number of planar trees with $n$ vertices is equal to the number
of planar binary trees with $2n-1$ vertices \cite{Conway}, there is a bijection
$\Psi$ between planar trees and planar binary trees.
If $t$ is a planar tree, the bijection can be realized 
recursively by, for example, $\Psi(\tunn)=\tunn$ and
\begin{eqnarray*}
\Psi(B_+(t))&=& \Psi(t)\vee\tunn,\\
\Psi(B_+(t_1,t_2,\dots,t_k))&=& \Psi(t_1)\vee \Psi(B_+(t_2,\dots,t_k)),
\end{eqnarray*}
where $B_+$ is the grafting operator defined by 
Kreimer \cite{Kreimer98,Kreimer}
to build rooted trees.
This bijection is also apparent in the existence of two methods
for the numerical solution of differential equations on Lie
groups: one based on planar trees \cite{Owren}, the other
on planar binary trees \cite{Iserles}.
Planar binary trees were chosen here because the
recursive formulas look simpler and because of the
mathematical results of Loday, Frabetti and collaborators.

\section{Acknowledgements}
I am very grateful to Dirk Kreimer and David Broadhurst 
for exciting discussions. I thank Alain Connes
for his suggestion of concentrating on the 
calculation of propagators. 
My warmest thanks go to Ale Frabetti and Jean-Louis
Loday for the wonderful day we spent together
talking about trees.
This is IPGP contribution \#xxxx.

\section{Appendix}
This appendix contains proofs of some of the statements
contained in the text.

\subsection{Proof of Eq.(\ref{sumlineq})}
Equation (\ref{sumlineq}) will be proved in two steps. 
Firstly, it will be shown that if $\phi^n(t)$ satisfies
Eq.(\ref{sumlineq}), then $\frac{\delta\phi^n(t)}{\delta v}=
\phi^{n+1}_\Sigma(t)$,
secondly, that the sum over trees is a solution of Eq.(\ref{lineq}).

The first step is to show that 
\begin{eqnarray}
\frac{\delta \phi^n(t;\{z\}_{1,n})}{\delta v(y)}=
\phi^{n+1}_\Sigma(t;y,\{z\}_{1,n}). \label{recure}
\end{eqnarray}
This will be proved inductively. According to the construction of
$\phi^n(\tunn)$ from the initial data $A$, the relation is true for
$t=\tunn$. Now we assume that the relation is true up to trees with
$2N-1$ vertices. Let $t$ be a tree with $2N+1$ vertices.
The functional derivative of Eq.(\ref{sumlineq}) gives us
\begin{eqnarray*}
\frac{\delta \phi^n(t;\{z\}_{1,n})}{\delta v(y)}&=&
\sum_{k=0}^n F\big(\frac{\delta \phi^k(t_1;\{z\}_{1,k})}{\delta v(y)},\\&&
  \hspace*{6mm}
  \phi^{n-k+1}_\Sigma(t_2;z,\{z\}_{k+1,n})\big) +\\&&
\sum_{k=0}^n F\big(\phi^k(t_1;\{z\}_{1,k}),\\&&
  \hspace*{6mm}
  \frac{\delta \phi^{n-k+1}_\Sigma(t_2;z,\{z\}_{k+1,n})}{\delta v(y)}\big).
\end{eqnarray*}
Since $t_1$ and $t_2$ have less vertices than $t$, relation (\ref{recure})
is true for them and we obtain
\begin{eqnarray}
\frac{\delta \phi^n(t;\{z\}_{1,n})}{\delta v(y)}&=&  
\sum_{k=0}^n F\big(\phi^{k+1}_\Sigma(t_1;y,\{z\}_{1,k}),\nonumber\\&&
  \hspace*{6mm}
  \phi^{n-k+1}_\Sigma(t_2;z,\{z\}_{k+1,n})\big) +\nonumber\\&&
\sum_{k=0}^n F\big(\phi^k(t_1;\{z\}_{1,k}),\nonumber\\&&
  \hspace*{6mm}
  \phi^{n-k+2}_{\Sigma\Sigma}(t_2;y,z,\{z\}_{k+1,n})\big), \label{deltaphi}
\end{eqnarray}
where $\phi^{k}_{\Sigma\Sigma}(t;\{z\}_{1,k})$
distributes the variables $z_1$ and $z_2$ over the $k$ positions 
  $z_1,\dots,z_k$
without changing the order of $z_3,\dots,z_k$. All ways to take
two number among $k$ are used, so it is clear that 
$\phi^{k}_{\Sigma\Sigma}(t;z_1,z_2,\dots,z_k)$ is symmetric
in $z_1,z_2$.

On the other hand, we know from Eq.(\ref{sumlineq}) that
\begin{eqnarray*}
\phi^{n+1}(t;y,\{z\}_{1,n}) &=& 
  F\big(\phi^0(t_1),
  \phi^{n+2}_\Sigma(t_2;z,y,\{z\}_{1,n})\big)+\\&&\hspace*{-2cm}
  F\big(\phi^{n+1}(t_1;y,\{z\}_{1,n}),\phi^1(t_2;z)\big)+\\&&\hspace*{-2cm}
   \sum_{k=1}^n
  F\big(\phi^k(t_1;y,\{z\}_{1,k-1}),
  \phi^{n-k+2}_\Sigma(t_2;z,\{z\}_{k,n})\big).
\end{eqnarray*}
If we symmetrize $y$ in $\phi^{n+1}$, we obtain
\begin{eqnarray*}
\phi^{n+1}_\Sigma(t;y,\{z\}_{1,n}) &=& 
  F\big(\phi^0(t_1),
  \phi^{n+2}_{\Sigma\Sigma}(t_2;z,y,\{z\}_{1,n})\big)+\\&&\hspace*{-3cm}
  F\big(\phi^{n+1}_\Sigma(t_1;y,\{z\}_{1,n}),\phi^1_\Sigma(t_2;z)\big)+
\\&&\hspace*{-3cm}
   \sum_{k=1}^n
  F\big(\phi^k_\Sigma(t_1;y,\{z\}_{1,k-1}),
  \phi^{n-k+2}_\Sigma(t_2;z,\{z\}_{k,n})\big)+
\\&&\hspace*{-3cm}
   \sum_{k=1}^n
  F\big(\phi^k(t_1;\{z\}_{1,k}),
  \phi^{n-k+2}_{\Sigma\Sigma}(t_2;z,y,\{z\}_{k+1,n})\big).
\end{eqnarray*}
Comparing this with Eq.(\ref{deltaphi}), we see that the two
expressions  are identical, and the property is proved for $t$.

If we denote by $X$ the (formal) sum $X=\sum_t \phi^0(t)$, then
from Eq.(\ref{recure}) we obtain 
$\frac{\delta X}{\delta v(z)}=\sum_t \phi^1(t;z)$.
Using Eq.(\ref{sumlineq}) we can write
\begin{eqnarray*}
X &=& \sum_t \phi^0(t)\\
&=& \phi^0(\tunn)+\sum_{t\not=\bullet} F(\phi^0(t_1),\phi^1(t_2;z)).
\end{eqnarray*}
At this point intervenes the essential property Eq.(\ref{dectree})
that each tree
different from $\tunn$ is generated in a unique way by the grafting
of two trees $t_1$ and $t_2$. The sum over $t_1$ and $t_2$, which
are the branches of $t$, can be replaced by an unrestricted sum
over all trees $t_1$ and $t_2$.
Thus
\begin{eqnarray*}
X &=& \phi^0(\tunn)+\sum_{t_1,t_2} F\big(\phi^0(t_1),\phi^1(t_2;z)\big)\\
&=& \phi^0(\tunn)+F\Big(\sum_{t_1} \phi^0(t_1),\sum_{t_2}\phi^1(t_2;z)\Big)\\
&=& A+F\big(X,\frac{\delta X}{\delta v(z)}\big).
\end{eqnarray*}

A small extension of the previous result is necessary to treat
the case of QED.
The Schwinger equation is now $X=A+AF(X,\frac{\delta X}{\delta v(z)})$
and the solution is $X=\sum_t \phi(t)$, where $\phi(\tunn)$ is 
the usual initial data and the recurrence relation becomes
\begin{eqnarray*}
\phi^n(t;\{z\}_{1,n})&=&
\sum_{k=0}^n \sum_{k'=0}^{n-k} \phi^k(\tunn;\{z\}_{1,k})\\&&\hspace*{-23mm}
F\big(\phi^{k'}(t_1;\{z\}_{k+1,k+k'}),
  \phi^{n-k-k'+1}_\Sigma(t_2;z,\{z\}_{k+k'+1,n})\big).
\end{eqnarray*}
The reasoning is the same as for the simple Schwinger equation.
We start by proving that
$\frac{\delta\phi^n(t)}{\delta v}=\phi^{n+1}_\Sigma(t)$.
This is done by using the recursive definition of $\phi$ to write
both sides in terms of $\phi(t_1)$ and $\phi(t_2)$.
Then we write $X=\sum_t \phi^0(t)$ and we show that, because
of the recurrence relation for $\phi$, it satisfies
$X=A+AF(X,\frac{\delta X}{\delta v(z)})$.

In this paper, the treatment was restricted to linear integral operators $F$. 
However, the method can treat other cases.
For instance, if $F$ depends on $X^2$, this
square can be, in some sense, ``linearized'' by the planar trees.
If $X=\sum_t \phi^0(t)$, then
\begin{eqnarray*}
X^2 &=& {(\sum_t \phi^0(t))}^2 = \sum_{t_1} 
  \sum_{t_2} \phi^0(t_1) \phi^0(t_2)\\
&=& \sum_t \psi^0(t),
\end{eqnarray*}
where $ \psi^0(t)$ (and more generally $ \psi^n(t)$) is defined by
\begin{eqnarray*}
\psi^n(\tunn) &=& 0\\
\psi^n(t) &=&  \sum_{k=0}^n\phi^k(t_1)\phi^{n-k}(t_2).
\end{eqnarray*}

\subsection{Proof of Eq.(\ref{compeq})}
Equation (\ref{compeq}) will be proved by two methods.
Both are fast and easy.
In the first method, we define a Schwinger equation
\begin{eqnarray}
{ X \choose Z}={A \choose 0} + H{ X \choose Y},
  \label{quoieq}
\end{eqnarray}
where
\begin{eqnarray*}
H{ X \choose Z}={F\big(X,\frac{\delta X}{\delta v(z)}\big) \choose
G\big(X+Z,\frac{\delta X}{\delta v(z)}+\frac{\delta Z}{\delta v(z)}\big)}.
\end{eqnarray*}

This is a Schwinger equation whose solution is given by
Eq.(\ref{sumlineq}), where the map is
\begin{eqnarray*}
{\phi \choose \psi}(t)=\sum_{k=0}^n
\left( \begin{array}{c}
 F\big(\phi^k(t_1),
        \phi_\Sigma^{n-k+1}(t_2)\big) \\
  G\big(\phi^k(t_1)+\psi^k(t_1),\\ \hspace*{1mm} \phi_\Sigma^{n-k+1}(t_2)
  +\psi_\Sigma^{n-k+1}(t_2)\big)
  \end{array} \right),
\end{eqnarray*}
which is Eq.(\ref{compeq}).

The upper component of Eq.(\ref{quoieq}) is the first Schwinger equation
$X=A+F(X,\frac{\delta X}{\delta v(z)})$.
If we write $Y=X+Z$, the lower component is
$Y=X+G(Y,\frac{\delta Y}{\delta v(z)})$. 
Therefore, $X+Z$ is the solution of the composition of
equations. The map corresponding to this solution
is $\chi(t)=\phi(t)+\psi(t)$.

The second proof can also be useful.
It starts by adding a parameter $s$ to the Schwinger equations:
$X(s)=A+sF(X(s),\frac{\delta X(s)}{\delta v(z)})$ and 
$Y(s)=X(s)+sG(Y(s),\frac{\delta Y(s)}{\delta v(z)})$.
We take the $n$-th derivative of $Y$ with respect to $s$,
and we write $Y^{(n)}$ for its value at $s=0$. The chain
rule gives
\begin{eqnarray*}
Y^{(n)}=X^{(n)}+n\sum_{k=0}^n {n-1\choose k} 
  G\big(Y^{(k)},\frac{\delta Y^{(n-k-1)}}{\delta v(z)}\big).
\end{eqnarray*}
It can be shown recursively that
\begin{eqnarray*}
Y^{(n)}=n!\sum_{|t|=n} \big(\phi^0(t)+\psi^0(t)\big),
\end{eqnarray*}
where $\psi$ satisfies Eq.(\ref{compeq}). 
The result follows by expanding $X(s)$ and $Y(s)$ as
power series of $s$ and taking its value for $s=1$.

\subsection{Proof of Eq.(\ref{inverse})}
Let $X(s)=1+\sum_{t\not=\tunn} s^{|t|} \phi(t)$.
Let us show that the series for $Y(s)=1-1/X(s)$
is given by $Y(s)=\sum_{t\not=\tunn} s^{|t|} \psi(t)$,
where $\psi(t)$ is defined by Eq.(\ref{inverse}).

The first step is to prove that, if $P(t)$ is defined
by Eq.(\ref{defP}), then, with an abuse of notation,
\begin{eqnarray}
P(Y_n)=\sum_{k=1}^{n-1} Y_k\otimes Y_{n-k}, \label{PYn}
\end{eqnarray}
or more precisely
\begin{eqnarray*}
\sum_{|t|=n}P(t)=\sum_{k=1}^{n-1} \Big( \sum_{|u|=k} u\Big) 
   \otimes \Big( \sum_{|v|=n-k} v\Big).
\end{eqnarray*}
As usually, this will be proved recursively. 
The property is true for $n=2$, because
\begin{eqnarray*}
P(\tquatredeuxnn)=0,\quad P(\tquatreunnn)=\ttroisnn\otimes\ttroisnn.
\end{eqnarray*}
If this is true up to $n$, then, from Eq.(\ref{dectree})
\begin{eqnarray*}
\sum_{|t|=n+1}P(t) &=& \sum_{k=0}^{n} \sum_{|t_1|=k} \sum_{|t_2|=n-k} 
  P(t_1\vee t_2).
\end{eqnarray*}
By definition (\ref{defP}),
\begin{eqnarray*}
\sum_{|t|=n+1}P(t) &=& \sum_{k=1}^{n} \sum_{|t_1|=k} \sum_{|t_2|=n-k} 
  (t_1\vee \tunn)\otimes t_2 +\\&&
  \sum_{k=2}^{n} \sum_{|t_1|=k} \sum_{|t_2|=n-k}
  \sum_{i=1}^{n(t_2)} (t_1\vee u_i)\otimes v_i.
\end{eqnarray*}
We use property (\ref{PYn}) in the right-hand side :
\begin{eqnarray*}
\sum_{|t|=n+1}P(t) &=& \sum_{k=1}^{n} \sum_{|t_1|=k} \sum_{|t_2|=n-k} 
  (t_1\vee \tunn)\otimes t_2 +\\&&
  \sum_{m=1}^{n-1} \sum_{k=m+1}^n \sum_{|t_1|=n-k}
  (t_1\vee Y_{k-m})\otimes Y_m.
\end{eqnarray*}
In the second term, we use Eq.(\ref{dectree}) to sum over
$k$ by
\begin{eqnarray*}
\sum_{k=m+1}^n Y_{n-k}\vee Y_{k-m}=Y_{n-m+1}-Y_{n-m}\vee\tunn.
\end{eqnarray*}
Therefore
\begin{eqnarray*}
\sum_{|t|=n+1}P(t)
 &=& \sum_{k=1}^{n} \sum_{|t_1|=k} \sum_{|t_2|=n-k} 
  (t_1\vee \tunn)\otimes t_2 +\\&&\hspace*{2mm}
  \sum_{m=1}^{n-1} \sum_{|t_1|=n-m+1} 
  \sum_{|t_2|=m} t_1\otimes t_2-\\&&\hspace*{2mm}
  \sum_{m=1}^{n-1} \sum_{|t_1|=n-m} (t_1\vee \tunn)\otimes t_2
  \\
 &=& \sum_{m=1}^{n} \sum_{|t_1|=n+1-m} \sum_{|t_2|=m} 
  t_1\otimes t_2.
\end{eqnarray*}

To complete the proof, we start from 
$X(s)=1+\sum_{t\not=\tunn} s^{|t|} \phi(t)$
and we define $Y(s)=\sum_{t\not=\tunn} s^{|t|} \psi(t)$,
where $\psi(t)$ is given by
$\psi(\tunn)=0$,
$\psi(t)=\phi(t)-\sum \phi(u_i)\otimes \psi(v_i)$. Thus
\begin{eqnarray*}
\sum_{|t|>1} s^{|t|}\psi(t)=\sum_{|t|>1} s^{|t|}\phi(t)-
  \sum_{|t|>1} s^{|t|} \sum_{i=1}^{n(t)} \phi(u_i) \psi(v_i).
\end{eqnarray*}
From Eq.(\ref{PYn}), we see that $|t|=|u_i|+|v_i|$
and 
\begin{eqnarray*}
\sum_{|t|>1} s^{|t|}\psi(t)&=&\sum_{|t|>1} s^{|t|}\phi(t)-\\&&
  \Big( \sum_{|u|>0} s^{|u|} \phi(u)\Big)
  \Big( \sum_{|v|>0} s^{|v|} \psi(v)\Big).
\end{eqnarray*}
From the definition of $X(s)$ and $Y(s)$ we deduce
\begin{eqnarray*}
Y(s)-\psi(\ttroisnn)=X(s)-\phi(\ttroisnn)-1
-(X(s)-1)Y(s).
\end{eqnarray*}
From the definition of $\psi(t)$ we get 
$\psi(\ttroisnn)=\phi(\ttroisnn)$, and $Y(s)$
satisfies the equation
\begin{eqnarray*}
X(s)-X(s)Y(s)=1,
\end{eqnarray*}
or $X(s)=1/(1-Y(s))$.
To simplify the notation, we have assumed $\phi(\tunn)=1$.
The general case is proved similarly and leads to
Eq.(\ref{inverse}).

\end{fmffile}

\end{document}